\documentclass[aps,pra,twocolumn,showpacs]{revtex4}
\usepackage{amsmath,graphicx,tabularx,multirow}
\usepackage{color,bbm,amssymb,times}
\usepackage{xcolor}

\newcommand{\be}{\begin{equation}}
\newcommand{\la}{\langle}
\newcommand{\ra}{\rangle}
\newcommand{\ee}{\end{equation}}

\newcommand{\ba}{\begin{array}}
\newcommand{\ea}{\end{array}}
\newcommand{\daga}{^{\dagger}}
\newcommand{\sm}{{\scriptscriptstyle (M)}}
\newcommand{\BINArev}[1]{{#1}}
\newcommand{\mgap}[1]{{#1}}
\begin{document}
\title{Sudden death and sudden birth of quantumness for a harmonic
oscillator interacting with a classical fluctuating environment}
\author{Jacopo Trapani}
\affiliation{Dipartimento di Fisica dell'Universit\`a 
degli Studi di Milano, I-20133 Milano, Italia}
\author{Matteo Bina}
\affiliation{Dipartimento di Fisica dell'Universit\`a 
degli Studi di Milano, I-20133 Milano, Italia}
\author{Sabrina Maniscalco}
\affiliation{Turku Centre for Quantum Physics, Department of Physics and 
Astronomy, University of Turku, FI-20014 Turun yliopisto, Finland}
\author{Matteo G. A. Paris}
\affiliation{Dipartimento di Fisica dell'Universit\`a 
degli Studi di Milano, I-20133 Milano, Italia}
\affiliation{CNISM, Udr Milano, I-20133 Milano, Italy.}
\email[]{matteo.paris@fisica.unimi.it}
\begin{abstract}
\BINArev{We address the dynamics of nonclassicality for a quantum system
interacting with a noisy fluctuating environment described by a
classical stochastic field.} As a paradigmatic example, we consider a
harmonic oscillator initially prepared in a maximally nonclassical
\BINArev{state}, e.g. a Fock number state or a Schr\"odinger cat-like
state, and then coupled to either resonant or non-resonant external
field.  Stochastic modeling allows us to describe the decoherence
dynamics without resorting to approximated quantum master equations, and
to introduce non-Markovian effects in a controlled way. A detailed
comparison among different nonclassicality criteria and a thorough
analysis of the decoherence time reveal a rich \BINArev{phenomenology}
whose main features may be summarized as follows: i) classical memory
effects increase the survival time of quantum coherence; ii) a detuning
between the natural frequency of the system and the central frequency of
the \BINArev{classical field} induces revivals of quantum coherence.
\end{abstract}
\pacs{03.65.Yz, 03.65.Ta, 03.65.Xp}
\date{\today}
\maketitle
\section{Introduction}
The environment-induced decoherence is the prevailing explanation for
the loss of nonclassicality of an open quantum system, being
responsible for the relaxation of the system to a statistical mixture of
classical-like states \cite{har06,zur91,paz93}. In this framework, a 
non-zero 
temperature environment
is usually described in terms of a quantized ensemble of simple physical
systems, e.g.  harmonic oscillators or spins, spanning a wide frequency
range and interacting with the quantum system of interest through a
suitable interaction Hamiltonian. A set of approximations, such as Born
and Markov approximations, is then exploited to obtain a differential
master equation describing the dissipative dynamics of the open quantum
system \cite{Kos72,Gor76,Lin76,Ban84,car93,bre02,wei08}.
\par
In a Markovian approach, the environment time-correlation
functions are assumed to decay istantaneously compared 
to the typical time-scale of the system, i.e. memory effects have
no influence on the system dynamics. In this context, 
a \BINArev{thoroughly} studied quantum \BINArev{system} is the single-mode quantum \BINArev{harmonic}
oscillator interacting with a bosonic bath of oscillators. 
For such an open system, the decoherence time, ruling the 
transition from the quantum to the classical regime, may be identified 
by different nonclassicality criteria, which have been widely
investigated
\cite{gla63,sud63,gla69,lee91,lut95,kly96,tak02,egg97,vog00,ric02,ken04,lew05,cas05,kie08}
and compared \cite{paa11}. Extensions to multimode systems  
\cite{ths09,mir10,sol10,bri11} have been analyzed, and the decoherence 
process has been addressed extensively \cite{S1,S2}.
Besides the fundamental interest, the analysis of the
quantum-to-classical transition has relevant applications in the field 
of quantum technology. In fact, the generation and detection of 
nonclassical states is often a prerequisite to generate entanglement 
and discord for quantum information purposes in all-optical setups 
\cite{joy97,opq00,kim02,wan02,oli11,fer12}. 
\par
The assumption of weak coupling between the system and its 
environment, i.e. the Born approximation, is valid for a wide 
class of systems. On the other hand, the Markov assumption, is violated 
in several situations
of interest, e.g. in biological, optical, or solid-state systems
\cite{man06,xtl10,reb11,chi12}, where a more detailed description of 
the environment, including the spectral structure and the inherent
memory effects, is required
\cite{liu11,smi11,pii08,smi13}.  In this regime, decoherence may be
less detrimental, and the dynamics may even induce {\em re-coherence}.
For this reason a great attention has been devoted to the study of 
the corresponding non-Markovian dynamics in different systems ranging 
from quantum optics to mechanical oscillators and harmonic lattices 
\cite{man07,and08,paz09,vas09,vas10, gal1,cor12,caz13,ven14}. 
Besides, there are evidences that non-Markovian open quantum 
systems \cite{bre09,riv10,lux10,vas11, man1} can be useful for quantum 
technology \cite{cry11,est11,man2}.  
\par
There are two main paradigms to describe the dynamics of open quantum
systems: on the one hand, as mentioned above, one may look at system and
enviroment as a single global quantum system whose evolution is governed
by an overall unitary \BINArev{operator}. Upon tracing out the
environment' degrees of freedom, we then obtain the dynamics of the
system.  On the other hand, we may consider the open quantum system
under the action of external random forces, i.e. coupled to a stochastic
classical field. Here the partial trace is substituted by the average
over the different realizations of the stochastic field.  While the
system-enviroment approach is more fundamental in nature, the
approximations employed to achieve manageable dynamical equations often
precludes a detailed description of the dynamics.
Indeed, systems of interest for quantum technology generally interact
with complex environments, with many degrees of freedom, and a fully 
quantum description may be challenging or even unfeasible. In these
situations, classical stochastic modeling of the environment 
represents a valid and reliable alternative.  In fact, it has been shown
that for certain system-environment interactions a classical description
can be found that is completely equivalent to the quantum description
\cite{hel09,hel11,cro14,sar13}. Besides, there are
various experimental evidences that many quantum systems of interest
interact with classical forms of noise, typically Gaussian noise
\cite{ast04,gal06,abe08}.
\par
In this paper, we consider the paradigmatic case of a quantum harmonic
oscillator coupled to a classical stochastic field (CSF)
\cite{gar83}. Here, the advantage of choosing a CSF description for the 
environment is twofold: on the one
hand stochastic modeling allows us to describe the decoherence dynamics
without resorting to approximated quantum master equations. On the other
hand, we may introduce non-Markovian effects in a
controlled way. For qubit systems, description of environment-induced 
decoherence by the interaction with classical fluctuacting field
has been successfully carried out
\cite{pal02,luk08,ben12,wol12,ben13,hun13,rev13,ben13a,ben14,ros15}.  
\par
We will assume that the harmonic oscillator is initially prepared 
in a maximally nonclassical state, e.g. a Fock number state or a 
superposition of (possibly mesoscopic) coherent states, the so-called 
{\em Schr\"odinger-cat} state, 
and perform a
detailed comparison of the decoherence times 
according to four different criteria for nonclassicality: the nonclassical 
depth \cite{lee91}, the negativity of the Wigner function \cite{hud74}
the Vogel criterion \cite{vog00}, based on the characteristic function, 
and the Klyshko criterion for the photon number distribution 
\cite{kly96}. While the sole nonclassical depth criterion represents 
a proper (i. e. necessary and sufficient) criterion for 
nonclassicality, the other quantities have the advantage of being 
good candidates for an experimental implementation. 
\par
Our results show that according to all the quantifiers 
of nonclassicality, the presence of time correlations (i.e. memory
effect) in the classical environment enhances the {\em survival time}
of, say, the Schr\"odinger cat state, i.e. it preserves coherence 
for a longer time compared to the Markovian case. Furthermore, these memory 
effects become more and more important as far as the central 
frequency of the stochastic 
field is detuned with respect to the natural frequency of the 
harmonic oscillator, up to inducing sudden death and sudden birth of
quantumness, i.e. collapse and revival of quantum coherence.
\par
The paper is organized as follows: in Sec. \ref{s:sys} 
we introduce the system under investigation and the 
stochastic modeling of the environment, as well as the details
of the system-environment interaction. We also describe the initial
preparation of the system, discuss their nonclassicality and introduce 
and all the figures of merit used in the subsequent Sections. In 
Section \ref{s:dq} we address in details the decoherence dynamics 
of the \BINArev{system} interacting with a classical environment described by 
the Ornstein-Uhlenbeck process.
We also evaluate the input-output fidelity of the corresponding quantum
channel and discuss its use as a potential indicator of 
non-Markovianity in our system.
In Section \ref{s:pwl}, we briefly analyze the decoherence 
dynamics for \BINArev{an} environment
described by a CSF with \BINArev{a power-law 
autocorrelation} function. 
Finally, Section \ref{s:out} closes the paper with some 
concluding remarks.
\section{The system}
\label{s:sys}
We consider a quantum harmonic oscillator interacting with a
classical external field. The Hamiltonian of the system may be
written as $H=H_0 + H_{\scriptscriptstyle SC}$ where the 
free \BINArev{and interaction 
Hamiltonians} are given by
\begin{align}
H_0 &=\hbar \omega_0 a^\dag a \\
H_{\scriptscriptstyle SC} &= \hbar \left [a \bar{B} (t) e^{i \omega t} + 
a\daga B(t) e^{-i \omega t} \right ]\,,
\end{align}
\BINArev{with $\omega_0$} the natural frequency of the 
oscillator \BINArev{and $B(t)$ a time-dependent 
fluctuating field with central frequency $\omega$ described by a
stochastic process with zero mean, whose complex conjugate is $\bar{B}
(t)$. From now on and throughout the paper, we will consider the
Hamiltonian $H$ rescaled in units of $\hbar \omega_0$. As a
straightforward consequence, the stochastic classical field $B(t)$, its
central frequency $\omega$ and the time $t$ become dimensionless
quantities (in units of $\omega_0$ and $\omega_0^{-1}$ respectively).} 
\par 
We assume that the system is
initially prepared in a Fock state 
$|n\rangle$ or in a superposition of coherent states with opposite 
phases, the so-called Schr\"odinger-cat state 
$ |\psi_{cat} \ra = \mathcal{N}^{-\frac12}
\big( |\alpha \ra + |-\alpha\ra \big)$
where $|\alpha\ra$ indicates a coherent state and the normalization 
constant is
$
\mathcal{N} = 2 \left[1+ \exp(-2|\alpha|^2)\right]\,.$
We focus on Fock or cat states since they have maximal nonclassical 
depth and thus represent the proper preparation to analyze the
quantum-to-classical transition in full details. Actually, as we will show in the next paragraph, any pure 
state other than Gaussian pure states would be equally good to 
address the dynamics of the nonclassical depth. On the other hand, 
sufficient criteria as the Vogel criterion and the Klyshko criterion 
do depend on the specific state under investigation, and thus having
in mind a specific class of states will be of help to properly
address the detection of nonclassicality in realistic conditions.
\par
The nonclassical depth $\eta$ of a quantum state \cite{lee91} is 
a quantitative measure of its nonclassicality, and is defined as the 
minimum number of photons to be added to a state in order to erase all 
of its quantum features. In terms of the $s$-ordered Wigner functions, the 
nonclassical depth is given by 
$$
\eta=\frac12 (1-\bar s)\,,
$$
where $\bar s$ is the largest value of $s$ for which the corresponding
$s$ ordered Wigner function $W_s[\rho](\alpha)$ is positive and may be
seen as a classical probability distribution.  In turn, we 
have $0\leq \eta\leq 1$.
The $s$-ordered characteristic function and the $s$-ordered Wigner 
function for the Fock states $|n\rangle$ and the cat states
$|\psi_{cat}\rangle$, as well as the cat's matrix elements in the Fock
basis, are given in Appendix A. As it is apparent from their expressions, 
the $s$-ordered Wigner functions of both classes of states are not positive function for any $-1<s 
\leq 1$. Correspondingly, the nonclassical depth $\eta$ of a Fock or cat 
state is equal to 
one \cite{tak02} independently on $\alpha$ or $n$,
i.e. the cat and the number states are maximally nonclassical 
states independently on their energy, as the first positive
Wigner function corresponds to $s=-1$, i.e. the 
Husimi Q function. More generally, we have that the nonclassical 
depth is $\eta = 1$ \cite{lut95} for any pure state other
than Gaussian pure states (squeezed coherent state); squeezed states have
$0\leq \eta \leq \frac12$ depending on the squeezing parameter, while
coherent state have $\eta=0$, properly capturing the fact that they are the
closest analog to classical states for the quantum harmonic oscillator.
\par
The Hamiltonian in the interaction picture reduces to:
\begin{eqnarray}
H_I (t) =  a e^{-i \delta t}\bar{B}(t) 
+ a\daga  e^{i \delta t}B(t) 
\end{eqnarray}
where $\delta=1-\omega$ is the detuning between the natural 
frequency of the oscillator and the central frequency of the CSF \BINArev{(in units of $\omega_0$)}.
The corresponding evolution operator is given by
\begin{align}
U(t) = {\cal T} \exp\left\{ - i \int_0^t \!\! ds\, H_I(s) \right\}\,,
\end{align}
where ${\cal T}$ denotes time ordering. Notice, however, that 
as far as $B(t_1) \bar B(t_2) = [B(t_1) \bar B(t_2)]^{*}$, 
the two-time commutator $[H_I(t_1),H_I(t_2)]$ 
is proportional to the identity
\be
\label{commutator}
[H_I(t_1),H_I(t_2)] = 2\, i \sin \left[\delta (t_2-t_1)\right]\, 
B(t_1) \bar B (t_2)\, \mathbb{I}\,,
\ee
and this form allows to evaluate time ordering using the 
Magnus expansion \cite{mag1,mag2}, which \BINArev{results} to be exact already at 
the second order. According to the Magnus expansion, the 
evolution operator may be written as
\be
U(t) = \exp (\Omega_1 + \Omega_2) 
\ee
where:
\begin{align}
\Omega_1 &= -i \int_0^t\!\! ds_1\, H_I (s_1) = a\daga \phi_t  - a \phi^{*}_t  \\
\phi_t &= - i \int_0^t\!\! ds_1\, e^{i\delta s_1} B(s_1) 
\end{align}
and
\begin{align}
\Omega_2 &=  \frac12 \int_0^t\!ds_1\!\int_0^{s_1}\!\! ds_2\,
\, [H_I(s_1),H_I(s_2)] \propto \mathbb{I}. 
\end{align}
Since $\Omega_2$ is proportional to the identity 
we may write the evolution of an initial density
operator $\rho(0)$ as
\be
\rho (t) = \left[e^{\Omega_1}\rho (0)\, e^{\Omega_1^*} \right]_B
= \left[ D(\phi_t) \rho (0) D^\dag (\phi_t)\right]_B
\label{avr}
\ee
where $D(\mu)=e^{\mu a^\dag - \bar\mu a}$ is the 
displacement operator and $[\,\dots]_B$ denotes the average over the
different realization of the stochastic process. 
Eq. (\ref{avr}) shows that the interaction Hamiltonian with a
classical field results in a time-dependent displacement 
 \BINArev{of argument $\phi_t$, related to the classical field $B(t)$ 
and, then, strongly affected by its stochasticity.}
\par
In our system we assume that the CSF 
$B(t)=B_x(t) + i B_y(t)$ is described by a Gaussian 
stochastic  \BINArev{process} with zero mean 
$[B_x(t)]_B = [B_y(t)]_B=0$ and \BINArev{diagonal structure of the autocorrelation 
matrix}
\begin{align}
\label{kernel}
\left[B_x (t_1) B_x (t_2)\right]_B & =\left[B_y (t_1) B_y (t_2)\right]_B =
K(t_1,t_2) \\ 
\left[B_x (t_1) B_y (t_2)\right]_B & =\left[B_y (t_1) B_x (t_2)\right]_B =
0,
\end{align}
\BINArev{with (dimensionless) kernel autocorrelation function $K(t_1,t_2)$.}
\par
Using the Glauber decomposition \cite{gla69} for the 
initial state 
\be
\label{cha}
\rho (0) =  \int\!\frac{d^2 \mu}{\pi}\, 
\chi_0 [\rho (0)] (\mu)\, D\daga (\mu)\,,
\ee
where the (\BINArev{symmetrically ordered}) characteristic function is given by $\chi_0 [\rho] 
(\mu) = \hbox{Tr}[\rho\, D(\mu)]$, we may write the evolved 
state as \be
\rho (t) =  \int\!\frac{d^2 \mu}{\pi}\, 
\left[e^{\mu \phi^{*} (t) - \mu^{*} \phi (t)}\right]_B 
\, \chi_0[\rho (0)] (\mu) D\daga (\mu)\,,
\ee
where, for any Gaussian stationary process, we may write 
\be
\left[e^{\mu \phi^{*} (t) - \mu^{*} \phi (t)} \right]_B = 
e^{-|\mu|^2 \sigma(t)}
\ee
\BINArev{and $\sigma(t)$ (following Ref.~\cite{puri}) can be} expressed as 
\begin{align}
\sigma(t) = \int_0^t\! \int_0^t\! ds_1 ds_2\,
\cos\left[\delta(s_1-s_2)\right]\,K(s_1, s_2)\,.
\label{sig}
\end{align}
The $s$-ordered characteristic function $\chi_s [\rho(t)] (\mu)$
of the evolved state is given by
\be
\label{charaevolution} 
\chi_s [\rho(t)] (\mu) = \chi_0 [\rho (0)] (\mu)\, e^{\frac12 |\mu|^2 
\left[ s -2\sigma(t)\right]}\,,
\ee
which corresponds to a Gaussian noise channel \cite{gla63,hol73}:
\be
\label{gausscha}
\rho (t)=G[\rho (0)] = \int\!\frac{d^2 \gamma }{\pi \sigma(t)} 
e^{-\frac{|\gamma|^2}{\sigma(t)}}\, D(\gamma )\rho (0)\, D\daga (\gamma), 
\ee
\BINArev{where $\sigma(t)$ in Eq. (\ref{sig}) plays the role of the variance of the Gaussian channel.}
\par
In order to obtain quantitative results we focus 
on Ornstein-Uhlenbeck (OU) 
process \cite{uhl30,gar83}, with autocorrelation function given by 
\begin{align}
K(t_1,t_2) = \frac12 \lambda\gamma \, e^{- \gamma |t_1-t_2|}\,.
\label{ouk}
\end{align}
The main conclusions of our analysis, however, are  
independent on the specific feature of the process, as far
as we consider classical fields described by stationary Gaussian 
 processes.
In Eq. (\ref{ouk}) $\lambda$ is a coupling constant 
and $\gamma$ is a memory parameter equal to the inverse of 
the characteristic time of the environment \BINArev{(in units of $\omega_0$)}.
As we will show in the following Sections the memory effects
associated to the interaction with a classical OU field
allows the initial state to preserve its nonclassicality
for times longer than those achieved with a Markovian environment.
As we will see, the smaller is $\gamma$, the longer 
is the  survival time of quantumness at fixed values of the detuning $\delta$. 
Conversely, for $\gamma \gg 1$, the survival time of the cat 
approaches the Markovian values \cite{paa11}. Indeed, 
for $\gamma \gg 1$ the autocorrelation function in Eq. (\ref{ouk}) 
approaches a Dirac delta function.
\par
For the OU process $\sigma(t)$ \BINArev{may be explicitly written as}
\begin{align}
\sigma(t) =& \frac{\lambda\gamma}{(\gamma^2+\delta^2)^2}
\Big\{
\delta^2  (1+ t \gamma )- \gamma^2 (1- \gamma t)\notag  \\ & + e^{-\gamma t}
\Big[(\gamma^2-\delta^2)\cos\delta t - 2\gamma\delta \sin\delta t 
\Big]\Big\}
%
\end{align}
leading to the following approximated expressions in some particular 
regimes:
\begin{flalign}
\label{andam}
\sigma(t) &\simeq \lambda t + \frac{\lambda}{\gamma}\,
e^{-\gamma t} \cos \delta t  & \gamma \gg 1 \\
\sigma(t) &\simeq \frac{\lambda \gamma}{\delta^2}(1-\cos\delta t) 
& \gamma \ll 1, \delta \gg 1\\
\label{andam3}\sigma(t) &\simeq \frac{\lambda \gamma t^2}{2}(1-\delta^2 t^2) 
& \gamma \ll 1, \delta \ll 1\,.
\end{flalign}
Overall, the interaction with a classical environment corresponds to a 
Gaussian channel with the time-dependent width $\sigma(t)$, which fully
characterizes the dynamics.
\par
Finally, we notice that the map in Eq.~(\ref{gausscha}) is a solution
of the standard Born-Markov quantum optical master equation 
\begin{eqnarray}
\label{meq}
\frac{d}{dt} \rho (t) &=&  \frac{\Gamma}{2} (N+1) [2 a \rho(t) a\daga - a\daga a
\rho(t) - \rho(t) a\daga a] \nonumber\\ &+&\frac{\Gamma}{2} N[2 a\daga \rho(t) a -
a a\daga \rho(t) - \rho(t) a a\daga]\,
\end{eqnarray}
in the limits $N\gg 1$ and $\Gamma t\ll1$\BINArev{, where $N$ is the number of thermal photons
in the environment and $\Gamma$ the dissipation rate}. 
Eq. (\ref{meq}) describes the open-system dynamics of a harmonic oscillator
(weakly) interacting with a Markovian bath of harmonic oscillators
at \BINArev{the temperature $[\log(1+N^{-1})]^{-1}$}.
In other words, in the regime of high temperature and short times, 
the interaction with a quantized environment is
equivalent to the interaction with a classical stochastic field. The explicit
mapping is provided by the relation $\sigma (t) 
\longleftrightarrow  \Gamma N t$. Further insight about the meaning of 
the involved parameters may be gained using a short-time, 
detuning-independent, approximation for $\sigma(t) \simeq
\frac 12 \lambda \gamma t^2$.
\section{Dynamics of quantumness}
\label{s:dq}
In this Section we address in details the quantum-to-classical
transition, according to four different criteria, for a Schr\"odinger 
cat and a Fock state interacting with a CSF. 
We evaluate the decoherence times and analyze whether and how 
these may increase for a channel with memory, compared to a 
Markovian one. We also discuss the role of detuning in producing
\BINArev{collapse and re-coherence effects (sudden death and birth of quantumness)}.
\subsection{Nonclassical depth}
\label{s:ncld}
As mentioned above, the nonclassical depth was introduced 
as the minimum number of thermal photons needed to 
erase the quantum features of a given state \cite{lee91}. 
In the phase space, 
the nonclassical depth enters as the minimum width of the Gaussian 
convolution needed to transform the (possibly 
singular) Glauber $P$ function  of a given state into a positive 
function. According to this measure, Fock number states and Schr\"odinger-cat 
states are maximally nonclassical states independently on their energy.
\par
Accordingly, the spirit of the nonclassical depth criterion is to find 
the smallest interaction time $t_Q$ such that the $P$-distribution of the 
evolved state becomes positive, i.e. the evolved state is a statistical 
mixture of  coherent states. 
Indeed, the nonclassical depth criterion well captures the intuition 
of decoherence as relaxation of the system into a statistical mixture 
of classical states. 
\par
As we will see, the interaction with the CSF turns 
the initial $P$ distribution into a positive function after a finite
interaction time $t_Q$. In addition, depending on the value
of the \BINArev{dimensionless} parameters $\lambda$, $\gamma$ and $\delta$ we may also
observe revivals of coherence (sudden birth of quantumness).
In order to determine these thresholds, one should consider the 
evolved state $\rho(t)$ and evaluate the time-dependent value 
of the nonclassical depth. Actually, it is sufficient to evaluate the 
nonclassical depth only for the initial state since Eq. (\ref{charaevolution}) 
shows that the normally-ordered characteristic function $\chi_1 [\rho(t)] (\mu)$
(which generates the $P$ distribution) corresponds to the $\tilde s$-ordered 
characteristic function of the initial cat state $\chi_{\tilde s}[\rho(0)](\mu)$, 
where $\tilde s = [1-2 \sigma(t)]$. As the nonclassical depth of the 
cat or the Fock states is
equal to one, the $P$ distribution becomes positive when it turns into 
a Husimi Q function, which corresponds to $\tilde s= -1$. This happens 
in a finite \BINArev{(dimensionless)} time $t_Q$ that is straightforwardly 
defined by
\be
\sigma (t_Q) = 1.
\ee
For values of $t$ such that $\sigma(t) >1$, the $P$ 
distribution is a positive function and the state is classical.
It is worth noticing that the nonclassical depth criterion only depends 
on $\sigma(t)$, which is indipendent on the initial state parameter 
$\alpha$ or $n$. More generally, for a state with initial nonclassical
depth $\eta_0$ the decoherene time $t_Q$ is given by the solution of the 
equation $\sigma (t_Q) = \eta_0$.
\par
\begin{figure}[h!]
\centering
\includegraphics[width=0.48\columnwidth]{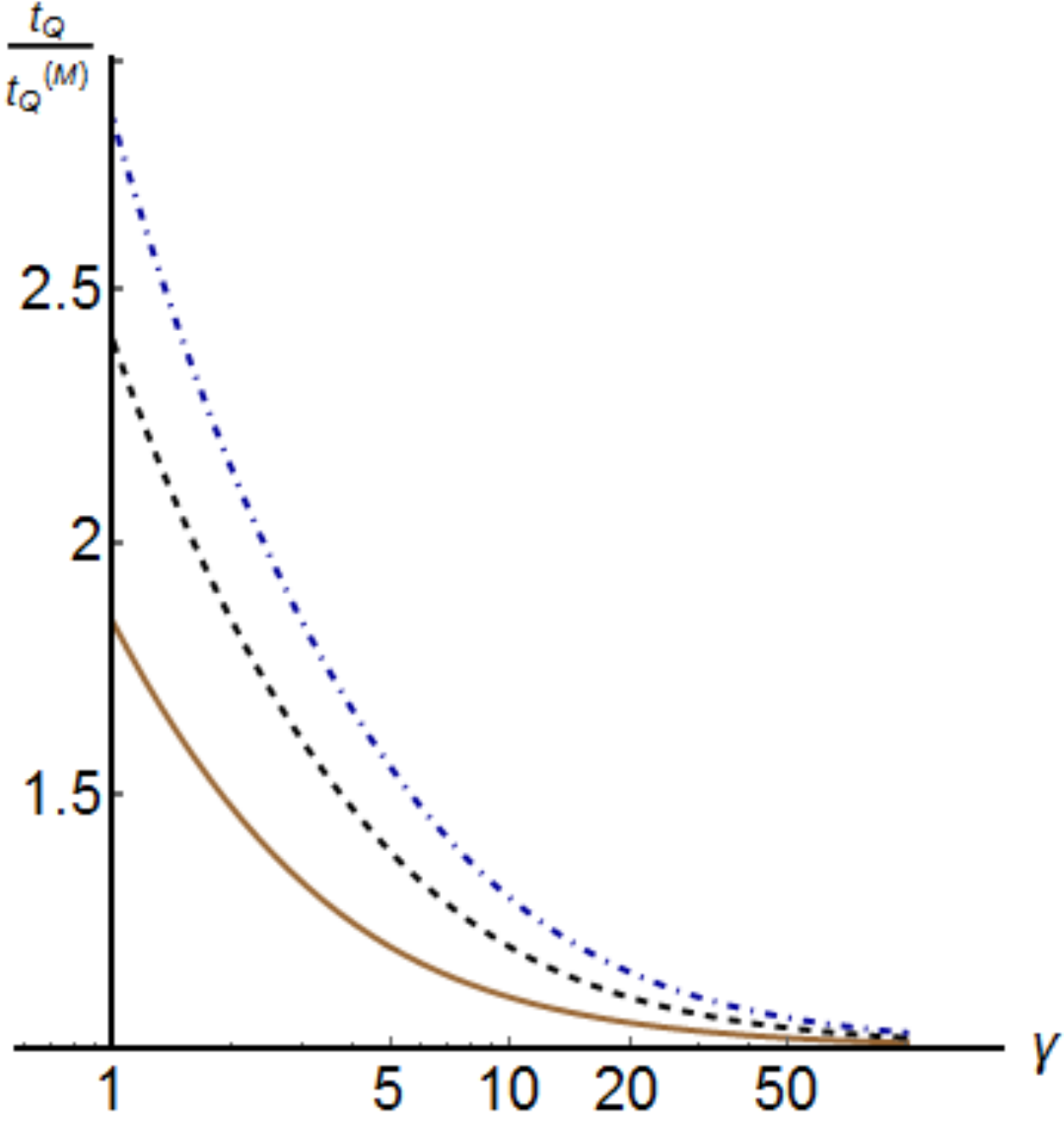} 
\includegraphics[width=0.51\columnwidth]{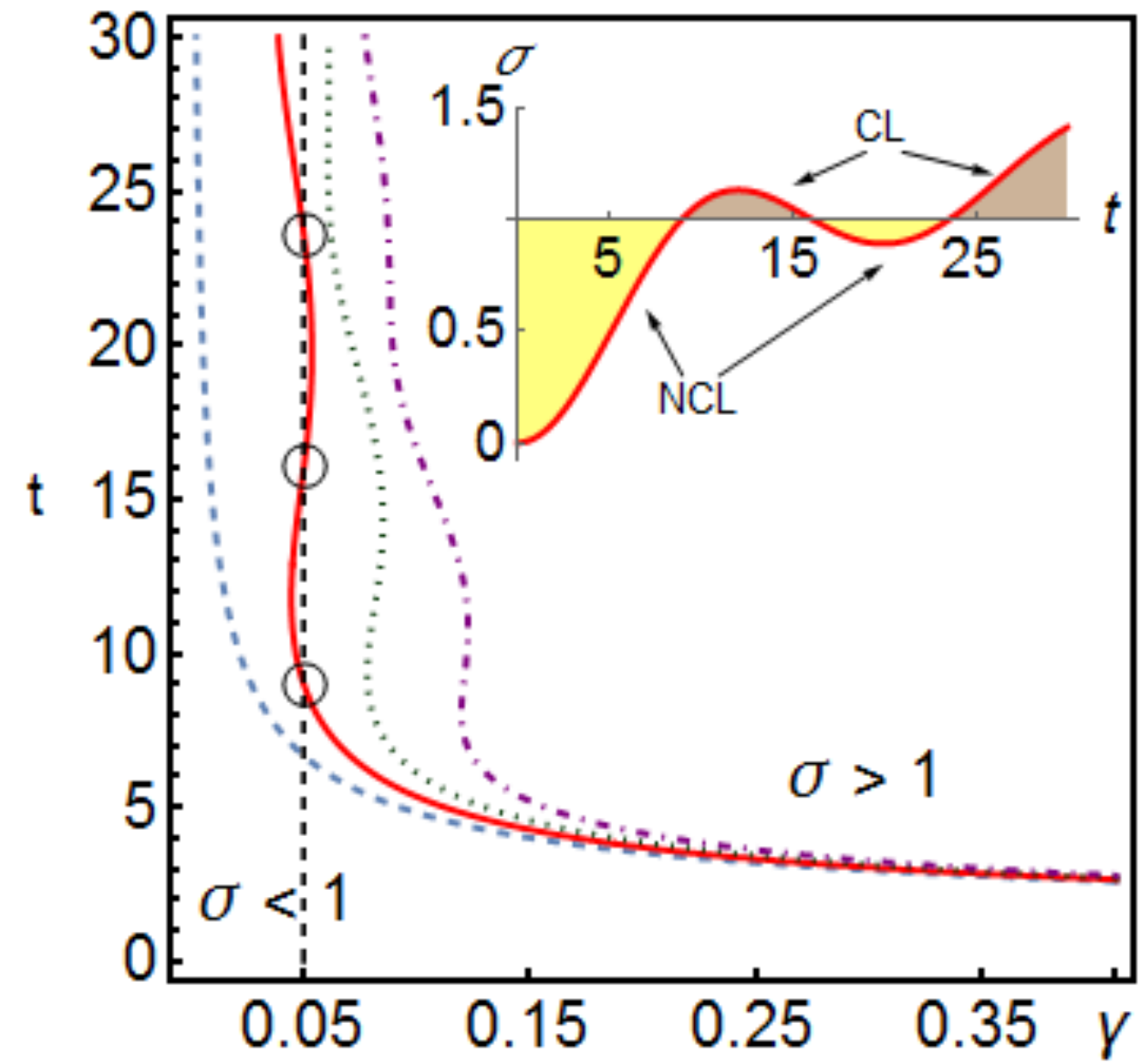}
\caption{(Color Online) Dynamics of quantumness according to the 
nonclassical depth criterion.  Left panel: dimensionless decoherence
time $t_Q$ for a resonant interaction as a function of the memory
parameter $\gamma$, for different values of the coupling $\lambda=1$
(solid brown), $\lambda=2$ (dashed black) and $\lambda=3$ (dot-dashed
blue).  For $\gamma \rightarrow \infty$, $t_Q$ approaches the Markovian
limit $t_Q^\sm$ independently on $\lambda$.  Right panel: contour plots
of $\sigma(\BINArev{t_Q})=1$, in the off-resonance case, as a function
of $\gamma$ for a fixed value of the coupling $\lambda=1$ and different
values of the detuning  $\delta=0.3$ (solid red), $\delta=0.4$ (dotted
green) and $\delta=0.5$ (dot-dashed purple).  The dashed blue
\BINArev{curve} is chosen as a reference for the resonant case
$\delta=0$.  In the regions lying to the left of the curves we have
$\sigma(t)<1$, i.e. nonclassicality.  The vertical line \BINArev{(dashed
black)} denotes points at fixed $\gamma = 0.05$ and the black circles
indicate the three solutions of $\sigma(\BINArev{t_Q})=1$ \BINArev{for
$\delta=0.3$}. \BINArev{Correspondingly, the regions of nonclassicality
(NCL) and classicality (CL) are highlighted in the inset}.
\label{f:fncld}
} \end{figure}

\BINArev{Let us firstly focus on} the resonant 
interaction ($\delta =0$). In this case $\sigma (t) $ reduces to:
\be
\sigma (t) = \lambda t + \frac{\lambda}{\gamma}
\left(e^{-\gamma t}-1\right)\,,
\ee
and the equation $\sigma(t)=1$ has a single solution for any 
pairs of values of $\lambda$ and $\gamma$. We 
thus have sudden death of quantumness without any revival.
As we anticipated in the previous Section, the autocorrelation
function of the process approches a Dirac delta in the limit 
of large $\gamma$. If we 
perform the limit at this stage we obtain
$\lim_{\gamma \rightarrow \infty} \sigma(t) = \lambda t$.
This form of $\sigma(t)$ coincides with that obtained using Eq. 
(\ref{meq}) and assuming that $\lambda = \Gamma N$. In other words, 
the limit $\gamma \gg 1$ leads to the Markovian regime. This also
confirms the idea that $\gamma$ plays the role of a memory 
parameter. More explicitly, its inverse set the time for which the 
field correlations cease to be significative. Large values of $\gamma$ 
describes environments with no memory of their previous configurations.
In the Markovian limit the decoherence time $t_Q^\sm$ is given 
by:
\be
t_Q^\sm = \frac{1}{\lambda} = \frac{1}{\Gamma N}.
\ee
In the present non-Markovian case, we have 
$$t_Q=
\frac{\gamma+\lambda}{\gamma\lambda} + \frac{1}{\gamma}
\,\BINArev{\xi\left ( -e^{-1-\gamma/\lambda}\right )}\,,$$
where \BINArev{$\xi(x)$} is the product-log function, i.e. the positive real
solution $y$ of the equation $x=ye^y$. Using this expression, it 
is possible 
to show numerically that $t_Q > t^\sm_Q$ for any value of $\gamma$
and $\lambda$, i.e. the non-Markovian character of the field
preserves the initial nonclassicality for longer times compared to 
the Markovian case. This is illustrated in the left panel of
Fig. \ref{f:fncld} where we show the ratio $t_Q/t_Q^\sm$ as a 
function of $\gamma$ for different values of $\lambda$: the
ratio is larger than unity for any value of $\gamma$ and it 
increases for increasing $\lambda$, i.e. nonclassicality is 
better preserved for larger coupling.  For increasing $\gamma$, 
the decoherence time $t_Q$ goes to the 
Markovian value independently on the value
of the coupling.
\par
Let us now analyze what happens if we turn on the detuning between 
the natural frequency of the system and the central frequency of 
the field. In this case the equation $\sigma(t)=1$ may have more than
one solution \BINArev{(fixing all the parameters $\delta, \gamma$ and $\lambda$)} and thus revivals of coherence may appear. 
In the right panel of Fig. \ref{f:fncld} we show the contour plots 
$\sigma(\BINArev{t_Q})=1$ as a function of time and $\gamma$ for different values
of the detuning $\delta$ and for a fixed value $\lambda=1$ of the
coupling. The regions lying
to the right of the curves correspond to $\sigma(t)>1$, i.e. 
classicality \BINArev{(CL), whereas regions of nonclassicality (NCL) $\sigma(t)<1$ lie to the left}.
There are two main effects: i) at fixed $\gamma$ the decoherence time $t_Q$ 
increases with the detuning, the effect is more pronounced for
smaller $\gamma$; ii) revivals of quantumness, i.e. sudden death
followed by sudden birth of quantumness, appear at fixed (and not too
large) values of $\gamma$.
This is illustrated in the right panel of Fig. \ref{f:fncld} \BINArev{and in the corresponding inset, where, for 
$\delta=0.3$ (solid red line) and $\gamma = 0.05$, $\sigma(t)$ displays re-coherence effects.}
Notice also that for increasing $\gamma$, revivals disappear and 
$t_Q$ becomes more and more independent on the detuning,
thus further confirming that for large $\gamma$ we are approaching the 
Markovian limit.
\subsection{Wigner negativity}
A different notion of nonclassicality is based on the negativity of
the Wigner \BINArev{function which is never singular, but it can take on negative values for 
nonclassical states, such as Fock states or superposition of 
coherent states\cite{ken04}}. The notion of nonclassicality arising from the 
negativity of the Wigner \BINArev{function} is not equivalent to the nonclassical depth 
and it has been linked to non-local properties \cite{coh97,ban98}.
Indeed, squeezed vacuum states display a positive Wigner function 
even though their nonclassical depth range from $\eta=0$ to
$\eta=\frac12$, increasing with energy. 
\par
We can evaluate the time $t_W$ in
which the $P$ function turns into a Wigner function in the very same way
we evaluated the nonclassical depth time in the previous section. The
condition that $t_W $ must satisfy, in order to change from a normally ordered
into a symmetrically ordered characteristic function, is
\be
\sigma (t_W) = \frac{1}{2}.
\ee
Exactly as the nonclassical depth criterion, the Wigner decoherence time 
depends only on $\sigma(t)$ and it is not affected by the initial state 
parameter $\alpha$ or $n$. For a state with initial nonclassical depth
equal to $\eta_0$, the Wigner decoherence time is the solution of 
$\sigma (t_W) = \eta_0 -1/2$ if $\eta_0 > \frac12$ or $t_W=0$ 
otherwise.
\par
In the Markovian limit $\gamma \gg 1$ the 
decoherence time $t_W^{(M)}$ of the cat or the Fock state 
is simply half of  $t_Q^{(M)}$
\be\label{tWM}
t_W^{(M)} = \frac{1}{\lambda} = \frac{1}{2 \Gamma N} =
\frac{1}{2}t_Q^{(M)}\,.
\ee
In the following, we are going to investigate whether the interaction 
with a stochastic field increases the coherence time of the cat
according to the Wigner negativity criterion, and to check how the 
relation in Eq. (\ref{tWM}) between $t_W$ and $t_Q$ is
affected by the memory parameter $\gamma$.  
\par
\begin{figure}[h!]
\centering
\includegraphics[width=0.48\columnwidth]{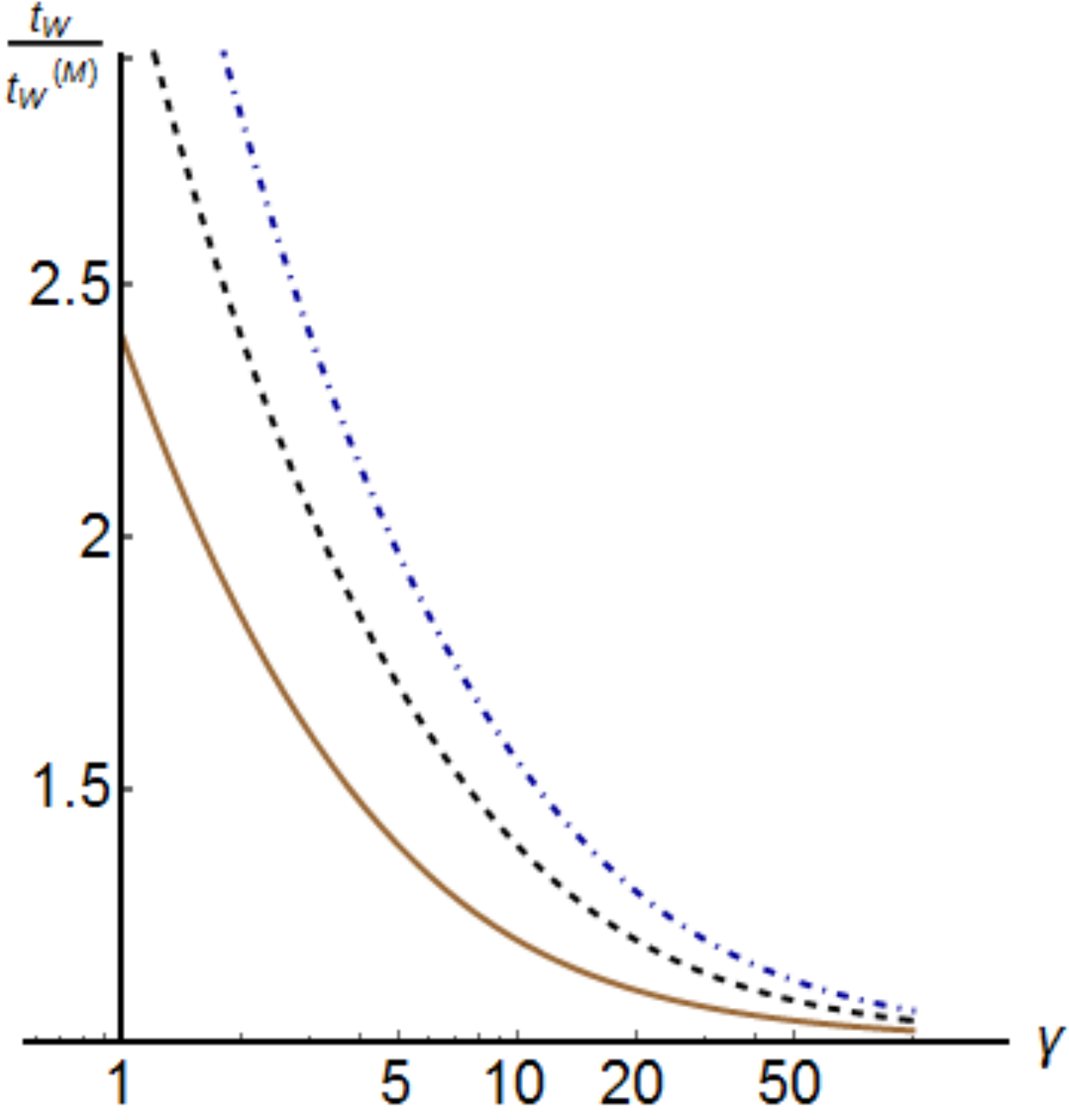} 
\includegraphics[width=0.51\columnwidth]{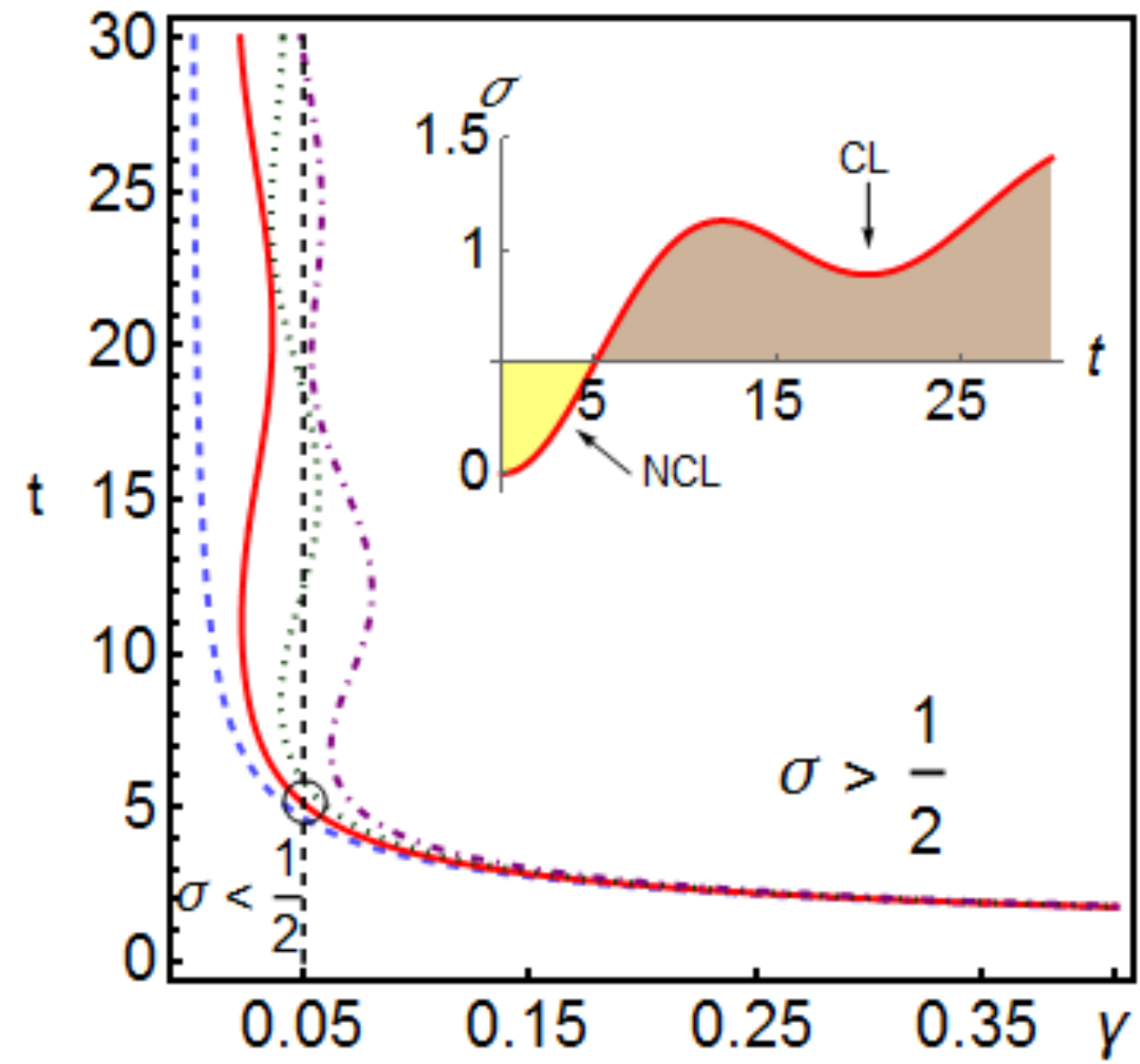}
\\ \vspace{0.4cm}
\includegraphics[width=0.9\columnwidth]{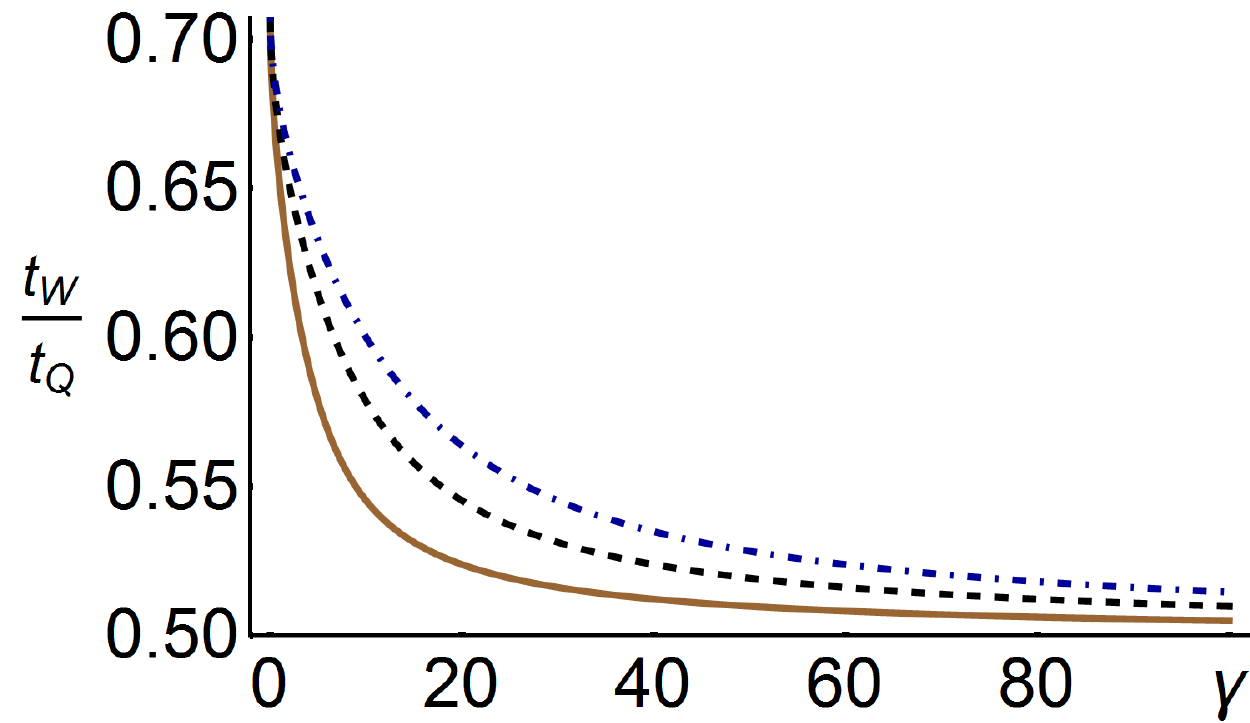}
\caption{\label{f:tW}
(Color Online) Dynamics of quantumness according to the Wigner
negativity criterion.  Upper left panel: Wigner decoherence time $t_W$
for a resonant interaction as a function of the memory parameter
$\gamma$, for different values of coupling $\lambda=1$ (solid brown),
$\lambda=2$ (dashed black) and $\lambda=3$ (dot-dashed blue).  For
$\gamma \rightarrow \infty$, $t_W$ approaches the Markovian limit
$t_W^\sm$ independently on $\lambda$.  Upper right panel: contour plots
of $\sigma(\BINArev{t_W})=\frac{1}{2}$, in the off-resonance case, as a
function of $\gamma$ for a fixed value of the coupling $\lambda=1$ and
different values of the detuning $\delta=0.3$ (solid red), $\delta=0.4$
(dotted green) and $\delta=0.5$ (dot-dashed purple).  The dashed blue
\BINArev{curve} is chosen as a reference for the resonant case
$\delta=0$.  In the regions lying to the left of the curves we have
$\sigma(t)<\frac{1}{2}$, i.e. nonclassicality.  \BINArev{The vertical
line (dashed black) denotes points at fixed $\gamma = 0.05$ and the
black circle indicates the  solutions of $\sigma(t_W)=\frac12$ for
$\delta=0.3$. Correspondingly, the regions of nonclassicality (NCL) and
classicality (CL) are highlighted in the inset.} Lower panel: the ratio
$t_W / t_Q$ as a function of $\gamma$ with same values of $\lambda$ as
in the upper left panel.  For $\gamma \gg 1$ the ratio approaches the
Markovian value $\frac{1}{2}$, whereas for $\gamma \ll 1$ it approaches
to the $\frac{1}{\sqrt{2}}$, due to the quadratic dependence on time of 
$\sigma(t)$. }
\end{figure}

The behaviour of the Wigner decoherence time is illustrated in Fig.
\ref{f:tW}. The upper left panel shows that $t_W$ is significantly
increased by the presence of time correlations in the CSF
\BINArev{(non-Markovian behavior)}, whereas the upper right panel
reveals \BINArev{re-coherence effects} for certain \BINArev{values of
the detuning and memory parameters}. In particular, the vertical black
line ($\gamma=0.05$) intercepts the solid red line ($\delta = 0.3$) just
once, which means that revivals \BINArev{of nonclassicality} displayed
in the nonclassical depth criterion (see Fig.1) \BINArev{are not
captured by} the Wigner criterion.  In the lower panel of Fig.
\ref{f:tW} we compare $t_Q$ and $t_W$ by showing their ratio as a
function of $\gamma$.  For large values of the memory parameter $\gamma$
(i.e. in the Markovian limit) the ratio approaches $\frac{1}{2}$,
according to Eq. (\ref{tWM}). In all the other cases, the ratio
increases and approaches the limiting value $\frac{1}{\sqrt{2}}$ for
$\gamma \rightarrow 0$. This may be understood as a consequence of the
behaviour of $\sigma(t)$, as reported in Eqs.  (\ref{andam}). Indeed,
$\sigma(t)$ is basically linear in time for large 
$\gamma$, whereas it shows a quadratic behaviour for $\gamma \ll1$.
\par
The study of the Wigner negativity criterion in the off-resonance regime
confirms the main conclusions we drew from the analysis of the nonclassical 
depth: for $\delta\neq 0$ the Schr\"odinger cat coherence survives
longer and sudden death and birth of nonclassicality appear, which is
expected \BINArev{by the analogy of the two considered criteria}.
\subsection{Vogel criterion}
According to the Vogel criterion \cite{vog00}, which establishes a
sufficient condition for nonclassicality, a state is nonclassical if
there exist some \BINArev{complex numbers $\mu=(u, v)$} such that the
normally ordered characteristic function satisfies 
\be
\label{vogel}
\left |\chi_1[\rho(t_V)] (\mu) \right | > 1,
\ee
\BINArev{where $ \chi_1[\rho(t)] (\mu) =\chi_0[\rho(t)] (\lambda){\rm e}^{e^{\frac12 |\mu|^2}}$.}
This is only a sufficient condition to characterize nonclassical states, 
but \BINArev{it has an advantage stemming from the fact that} the symmetric characteristic function can be
directly measured via balanced homodyne detection. The Vogel criterion
is then suitable for an experimental implementation \cite{lvo02}.
\par 
It is worth noticing, however, that 
in contrast with the two criteria shown 
previously, the Vogel criterion do depend on the state under
investigation, i.e. the smallest interaction time $t_V$ 
for which Eq. (\ref{vogel}) is satisfied, depends on the amplitude 
$\alpha$ for the Schr\" odinger state or on the specific Fock state $|n\ra$. 
Here, we consider cat states with real amplitude $\alpha = 
\alpha^{*}= \sqrt{2}$,  the reason of this choice being justified 
later (see Section D). The Fock state $|n=2\ra$ is chosen such that
the number of photons approximates the cat mean number of 
photons $\la a\daga a \ra \simeq 2$.
\par
\begin{figure}[h!]
\centering
\includegraphics[width=0.495\columnwidth]{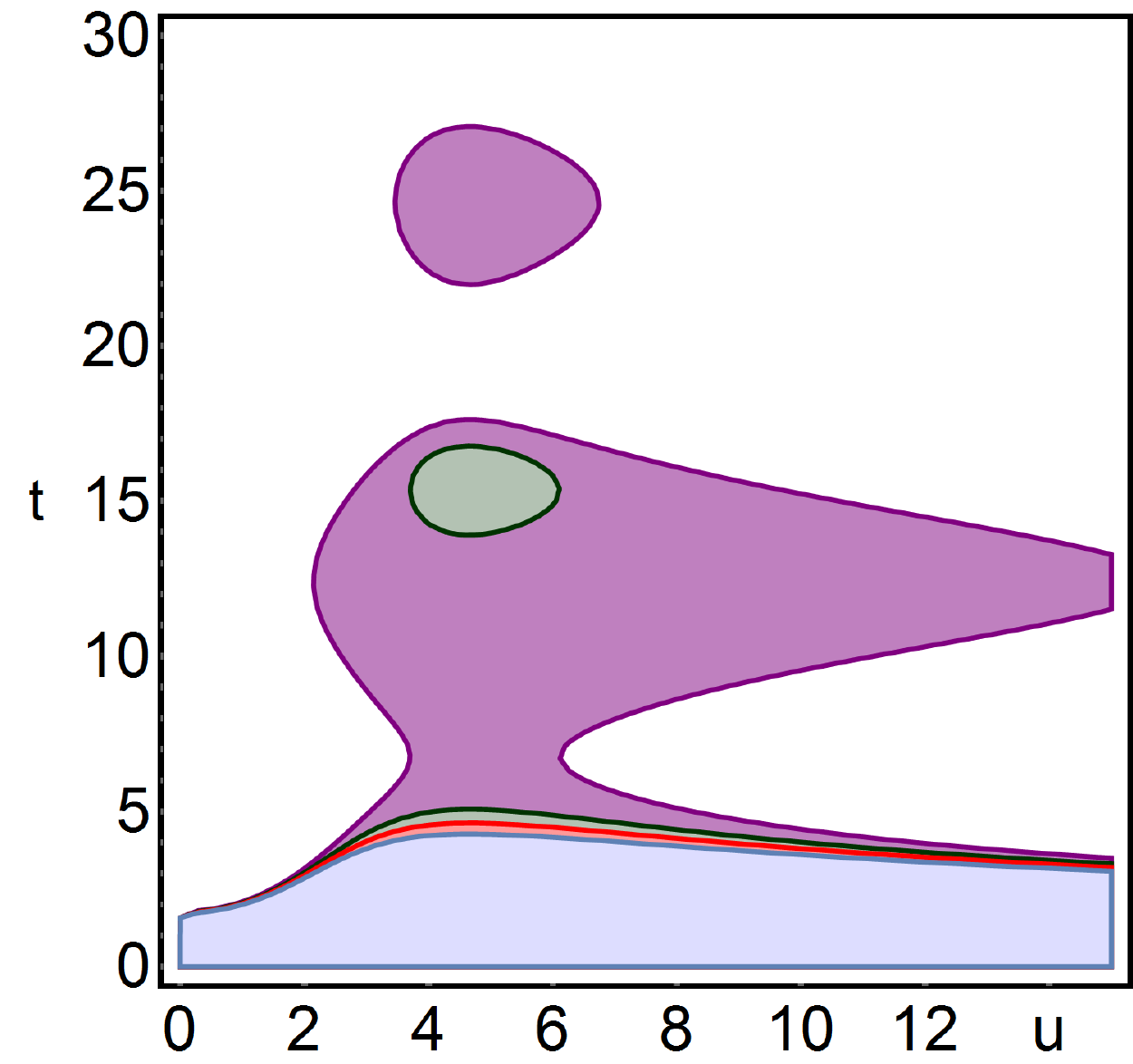}
\includegraphics[width=0.495\columnwidth]{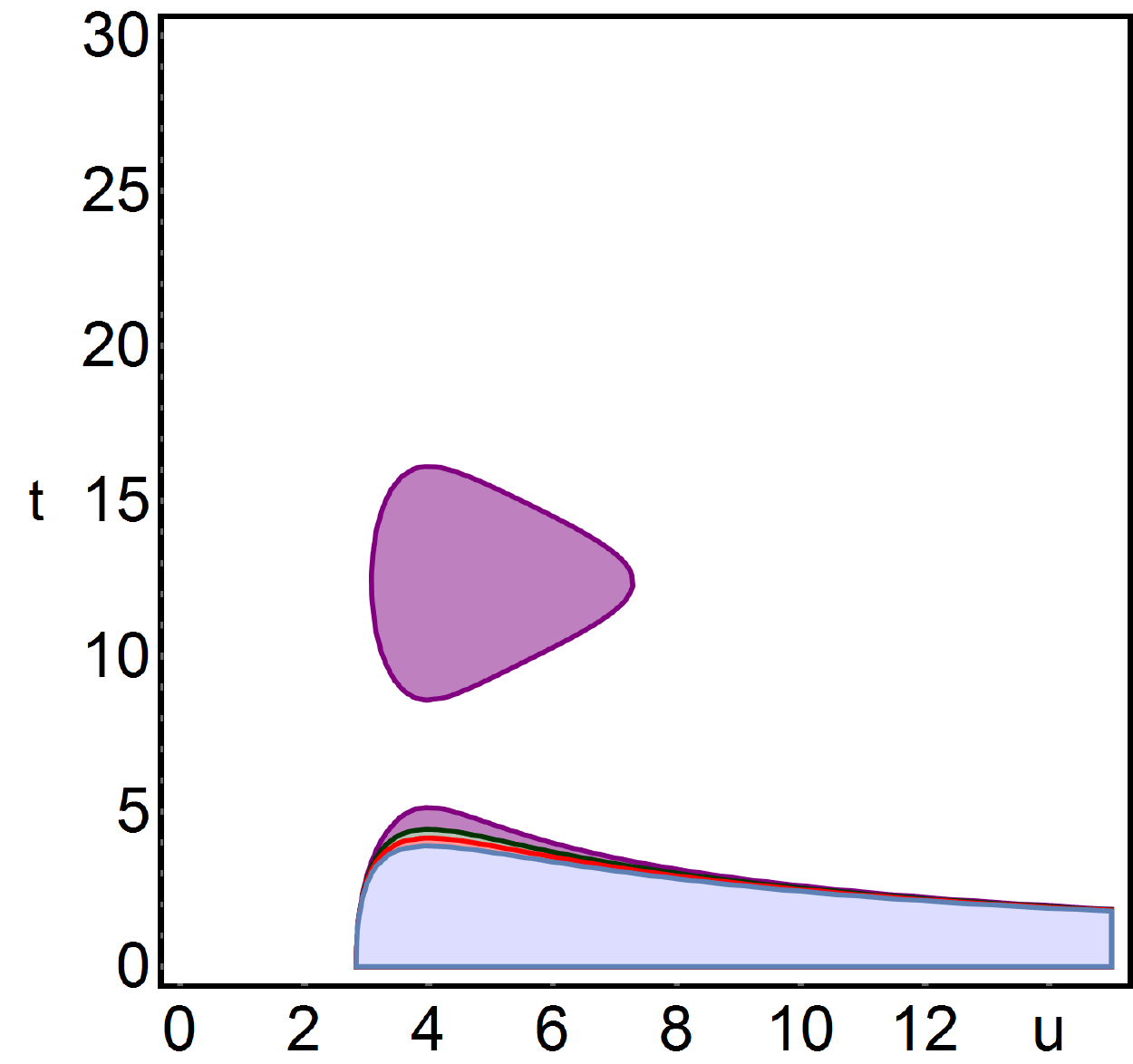}
\caption{\label{f:Vfig}(Color Online) Left panel: cat 
state Vogel time $t_V$ as a function of $u$, with $|\alpha| = \sqrt{2}$.
Right panel: Vogel time $t_V$ as a function of $u$ for Fock state
$|2\ra$.  In both panels, \BINArev{$\gamma = 0.05$ and} filled regions
correspond to $|\chi_1[\rho(t_V) (u,0)| >1$. From bottom to top: the
blue region represents the resonant interaction ($\delta=0$), whereas
the red ($\delta=0.3$), green ($\delta =0.4$) and  purple ($\delta
=0.5$) regions correspond to the off-resonance case. The spots for the
green and the purple regions indicate the presence of revivals
\BINArev{of nonclassicality.} } 
\end{figure}

\BINArev{The plots in the left and right panels of Fig. \ref{f:Vfig}, for cat and Fock states respectively, show the regions 
for which $ | \chi_1[\rho(t_V)] (\mu)|>1$}, as a function of 
${\rm Re}(\mu)= u$ \BINArev{(with $v=0$) and varying the detuning parameter $\delta$ (different colors)}. 
As it is possible to see in both figures, after a certain time $t_V$
nonclassicality disappears, but the sudden birth and sudden death of
quatumness is present \BINArev{also according to the Vogel criterion
(look, for example, at the green and purple regions)} and consistently
with the two previous criteria, as far as the off-resonance interaction
\BINArev{($\delta\neq 0$)} between the system and the CSF is set. 
\subsection{Klyshko Criterion}
\begin{figure}[h!]
\centering
\includegraphics[width=0.495 \columnwidth]{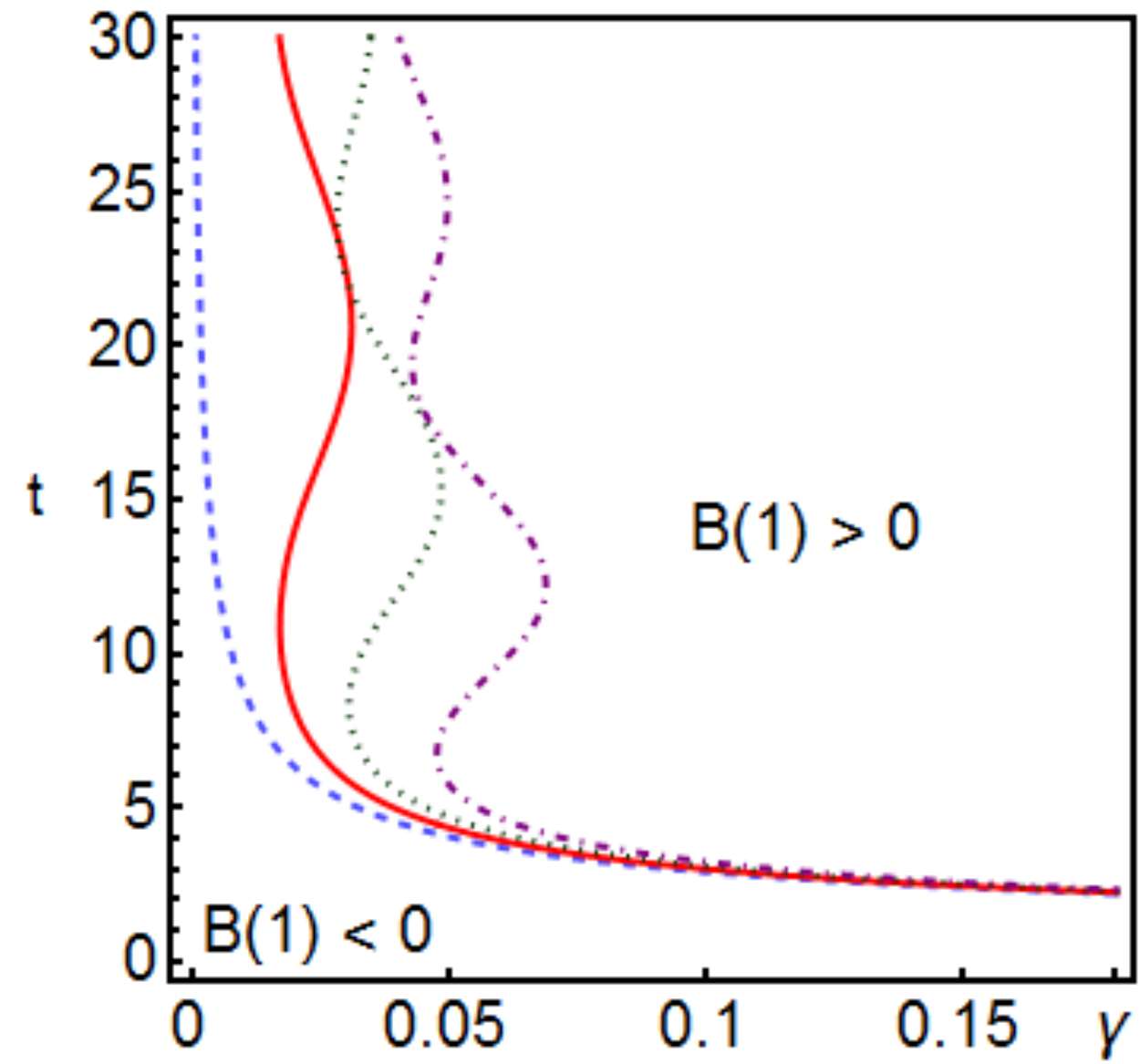}
\includegraphics[width=0.495 \columnwidth]{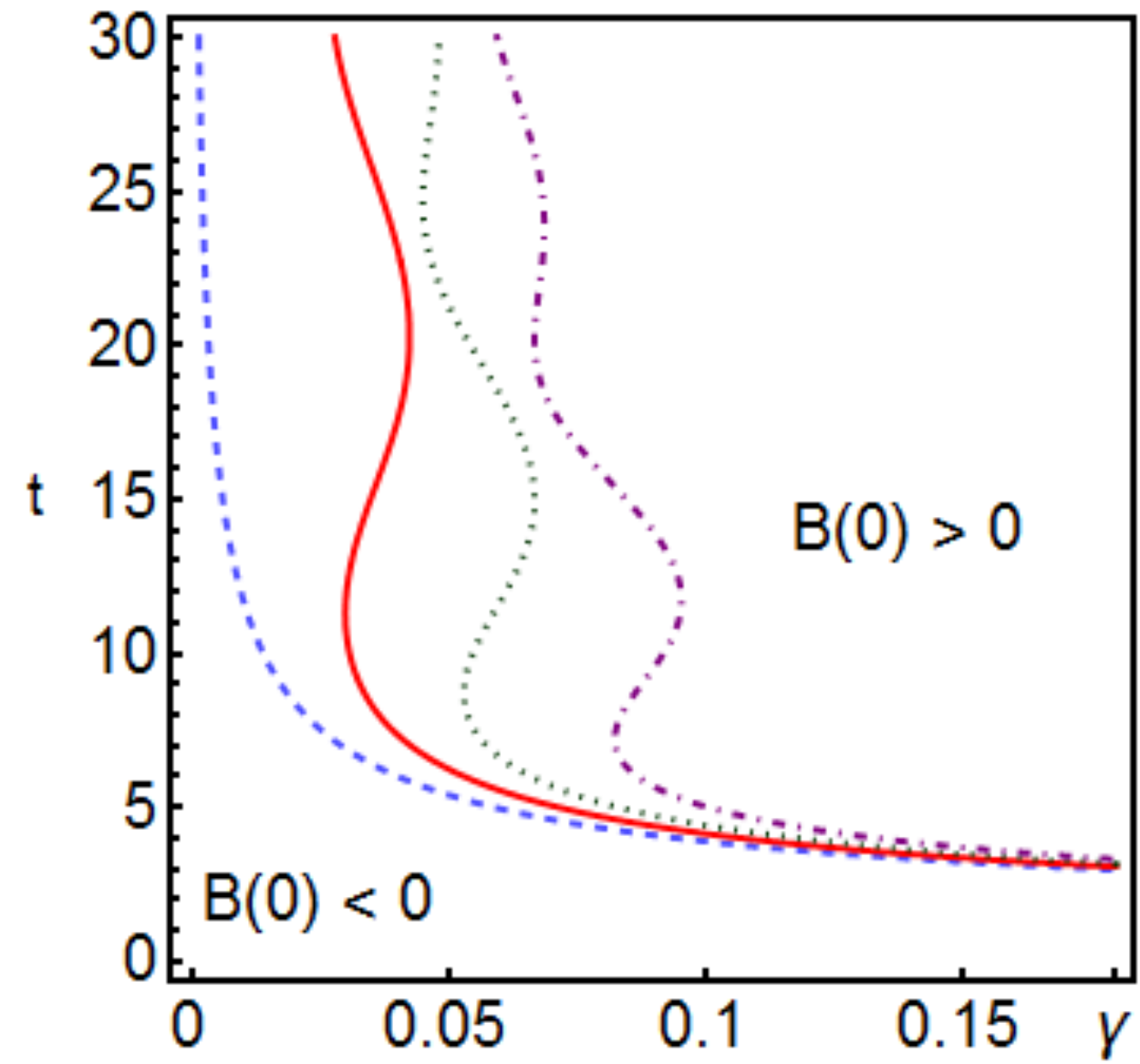}
\caption{\label{f:Kfig}
(Color Online) Left panel: \BINArev{Dimensionless} decoherence time
$t_K$ for the Klyshko criterion as a function of $\gamma$ for the cat
state with $\alpha = \sqrt{2}$.  Right panel: Decoherence time $t_K$ for
the Klyshko criterion as a function of $\gamma$ for the Fock state
$|2\ra$.  In both panels, \BINArev{dashed} blue curve represents the
resonant interaction ($\delta=0$), whereas solid red ($\delta=0.3$),
dashed green ($\delta =0.4$) and dot-dashed purple  ($\delta =0.5$)
curves refer to the off-resonance case. In the regions lying to the left
of the curves we have $B(1)<1$, i.e. nonclassicality.  } 
\end{figure}

In 1996, Klyshko introduced another criterion for nonclassicality, which
is only sufficient \cite{kly96}, stating that if there exists an integer
number $n$ such that: 
\be \label{klyshko} B(n) = (n+2) p(n) p (n+2) -
(n+1) [p(n+1)]^2 <0, 
\ee 
where $p(n) = \la n | \rho\BINArev{(t)} | n \ra$ is the
photon number probability, then the state $\rho$ is nonclassical.
Exactly as for the Vogel criterion, this nonclassicality witness is
experimentally accessible as it is based on photon counting
measurements. In our analysis of the Schr\"odinger cat nonclassicality,
according to the Klyshko criterion, we found out that $B(1)$ becomes
negative after a certain time $t_K$ dependent on the detuning $\delta$
and the memory parameter $\gamma$. As it is shown in the left panel of
Fig. \ref{f:Kfig}, the Klyshko criterion confirms that the cat survival
time increases for short $\gamma$ and it is affected by detuning.
Also in this case, sudden death and birth of quantumness can be
observed, as for fixed $\gamma$ there exist more than one time $t_K$
that satisfies the Klyshko criterion ($\ref{klyshko}$). A similar 
behaviour is shown for the Fock state $|2\rangle$ in the right panel 
of Fig. \ref{f:Kfig}, the only difference being the use of the quantity 
$B(0)$ instead of $B(1)$ to detect the quantum-to-classical transition.
As we mentioned earlier, we have chosen $|\alpha|=\sqrt{2}$ for the cat. 
In turn, this choice maximizes the effectiveness of Klyshko criterion, 
i.e. is the value corresponding to the longest survival time 
by Klyshko criterion \cite{paa11}.
\subsection{A remark about decoherence times}
In the previous Sections we went through a quantitative analysis of the
nonclassicality dynamics of the Schr\"odinger cat and the Fock state,
analyzing four different nonclassicality criteria. We described how the
interaction \BINArev{of a quantized harmonic oscillator} with a CSF, in
terms of an OU process, allows \BINArev{to preserve the nonclassicality
of each input state for certain periods of times}  and this result has
been confirmed by each nonclassicality criteria. A quantitative analysis
for both input states is shown in Table \ref{tab:1}, where we report the
times corresponding to the sudden death of quantumness achieved
according to the four considered criteria, for several values of the
detuning $\delta$. In particular, they are obtained by fixing the value
of the parameter $\gamma=0.05$, which is responsible of an appreciable
memory effect in the considered OU process.  
\begin{table}[h!]
\BINArev{Schr\"odinger-cat state:} \quad\begin{tabular}{|l ||c| c| c| c|}
$\delta$&0 & 0.3 & 0.4 & 0.5   \\ \hline
$t_Q$ & 6.676 &8.982 & 47.467 &81.091\\
$t_W$ & 4.645 & 5.118 & 5.823 & 29.355\\
$t_V$ & 4.272 & 4.624 & 5.067 & 16.773 \\ 
$t_K$ & 4.054 & 4.349 & 4.694 & 17.700
\end{tabular}
\\ \vspace{0.2cm}
\BINArev{Fock (number) state:}\quad\hspace{1.2mm}
\begin{tabular}{|l ||c| c| c| c|}
$\delta$&0 & 0.3 & 0.4 & 0.5   \\ \hline
$t_Q$ & 6.676 &8.982 & 47.467 &81.091\\
$t_W$ & 4.645 & 5.118 & 5.823 & 29.355\\
$t_V$ & 3.886 & 4.140 & 4.425 & 5.128 \\ 
$t_K$ & 5.412 & 6.253 & 21.329 & 49.527
\end{tabular}
\caption{\label{tab:1} 
\BINArev{Dimensionless} decoherence times, obtained for $\gamma=0.05$, 
$\lambda=1$ and different values of the detuning $\delta$, 
corresponding to the sudden death of quantumness of 
the evolved Schr\"odinger
cat state (Upper Table) and the evolved Fock state (Lower Table), according to the four nonclassicality criteria: 
nonclassical depth ($t_Q$), Wigner negativity ($t_W$), 
Vogel criterion ($t_V$) and Klyshko criterion ($t_K$).}
\end{table}
\par
We notice that the times estimated with the Vogel
criterion (or the Klyshko criterion) are always shorter than the
nonclassical depth and the Wigner negativity decoherence times. This
is consistent with the fact that Vogel and Klyshko
criteria provide only sufficient conditions for the loss of
quantumness. Indeed, it is possible to still have an amount of
nonclassicality in the evolving state which is undetected by these two
criteria.  Actually, Diosi demonstrated that for some nonclassical
states the Vogel criterion is not satisfied \cite{dio00}. In other
words, the evolved cat \BINArev{or Fock} state may still show some quantumness,
according to other nonclassicality criteria, while the Vogel criterion 
is no longer violated. 
\subsection{Input-output fidelity}
The presence of oscillations in the dynamics of nonclassicality suggests
that some form of information backflow from the environment to the
system is taking place. This phenomenon is usually associated to quantum
non-Markovianity and we want to explore this connection, at least in a
qualitative way.  \mgap{In fact, the markovian character of the quantum map
$(\ref{gausscha})$ for a coherent input state may be easily proved
\cite{vas11}, but Markovianity on coherent states does not necessarily
imply Markovianity on Fock states or superposition of coherent states. 
\par
The non-Gaussian character of the channel under investigation 
prevents the analytic evaluation of non Markovianity 
measures based on fidelity \cite{vas11} or the Fisher information
\cite{lux10}.} On the other hand, since the dynamics induced by the 
interaction with the CSF is fully described by the 
quantum channel $(\ref{gausscha})$, the input-output fidelity, 
assessing the dissimilarity between the input state
and the output state of a quantum map, may be 
evaluated in a straightforward way as
\be
F_{IO}= \la \psi | \mathcal{E}(\rho) | \psi \ra\,,
\ee
where $\psi$ is the initial state, assumed to be a pure state.
\par
Actually, a non monotonous time evolution of $F_{IO}$ cannot be, in
general, interpreted as a signature of backflow of information from the 
environment to the system (one may construct examples where a system
interact with a markovian environment and still the IO fidelity
oscillates due to some unitary terms in the interaction Hamiltonian). 
On the other hand, we found that for our system $F_{IO}$ provides useful
information which may be relevant in the qualitative and quantitative 
characterization of non-Markovianity. 
\par
\begin{figure}[t]
\centering
\includegraphics[width=0.495 \columnwidth]{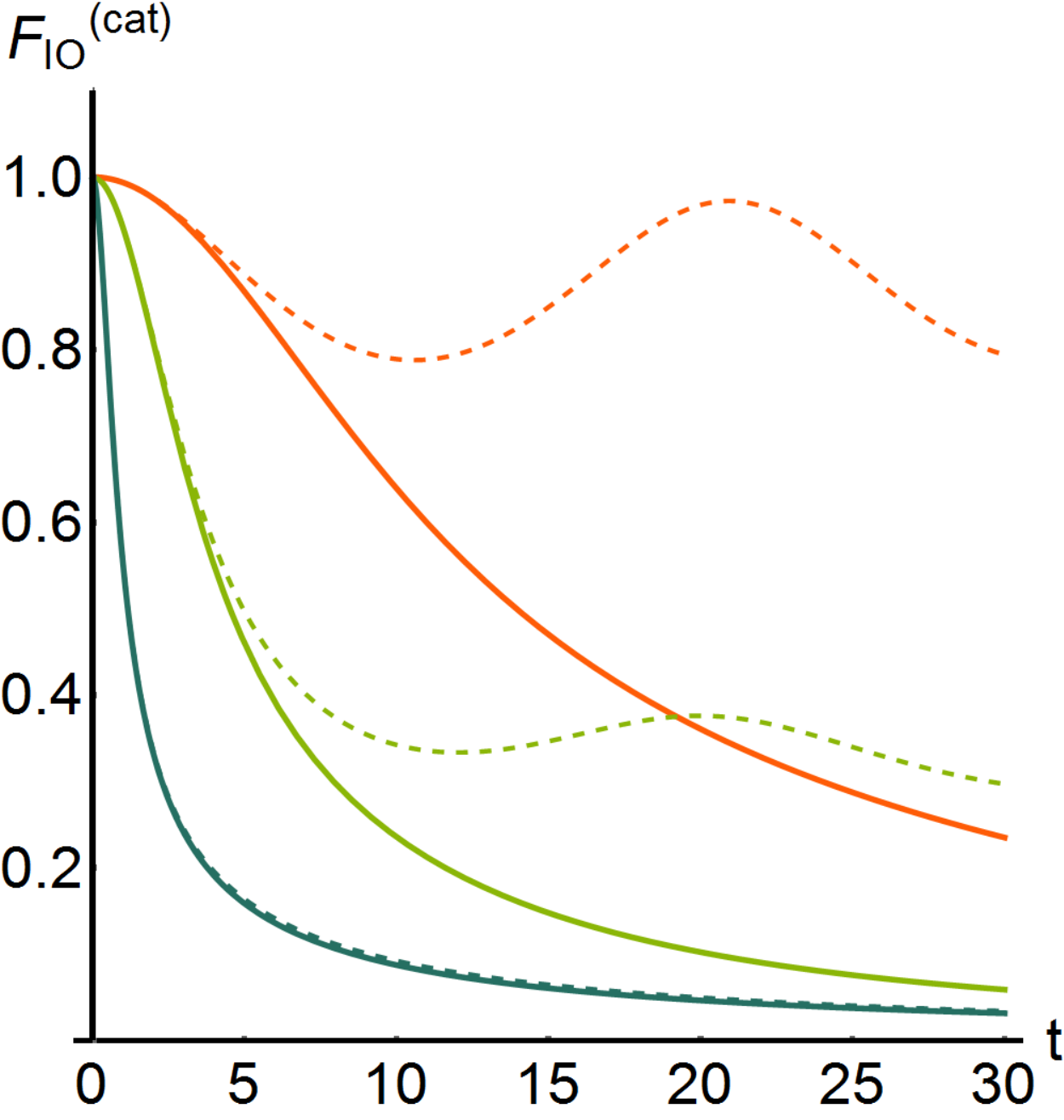}
\includegraphics[width=0.495 \columnwidth]{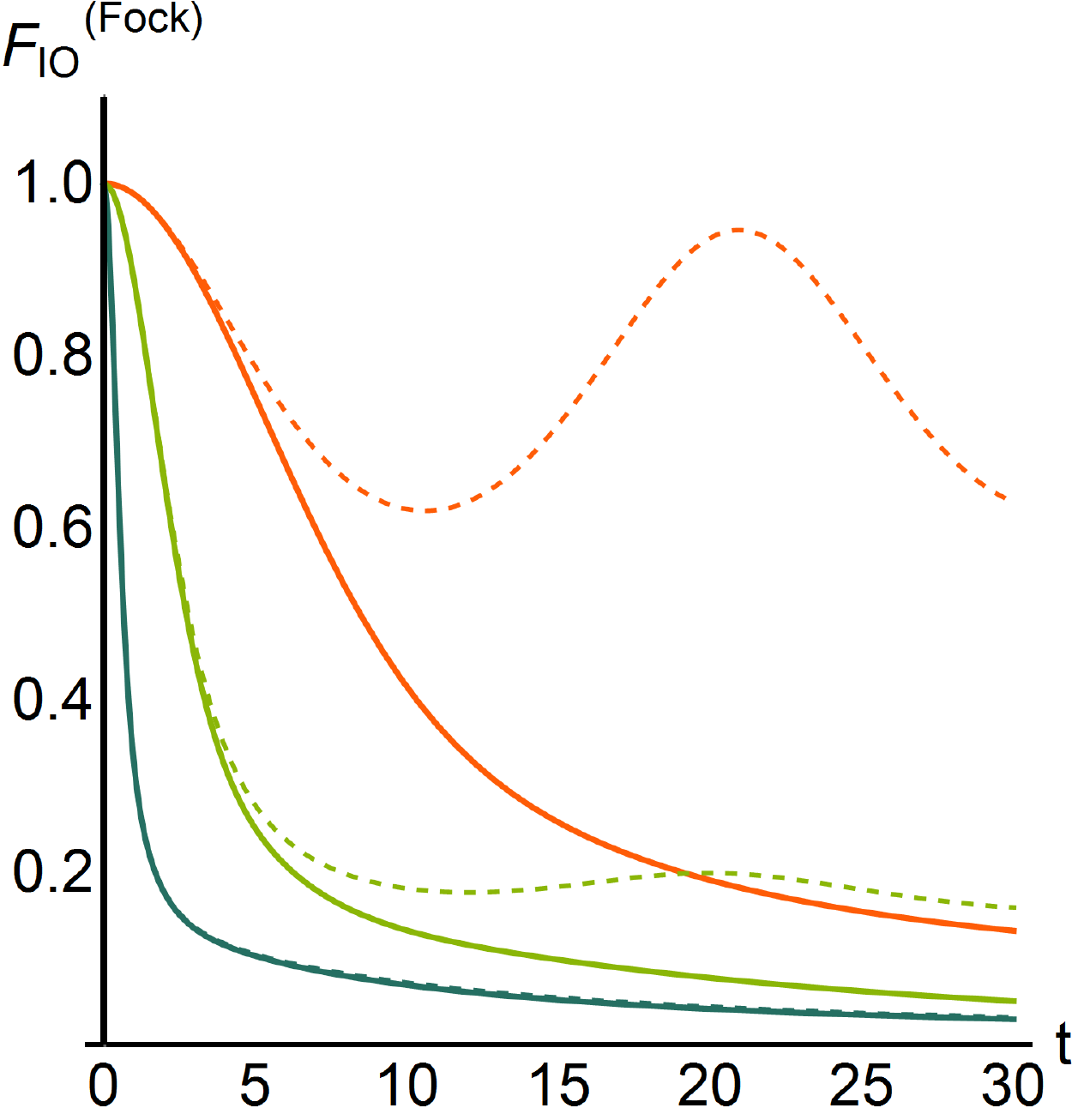}
\\ \vspace{0.2cm}
\includegraphics[width=0.9 \columnwidth]{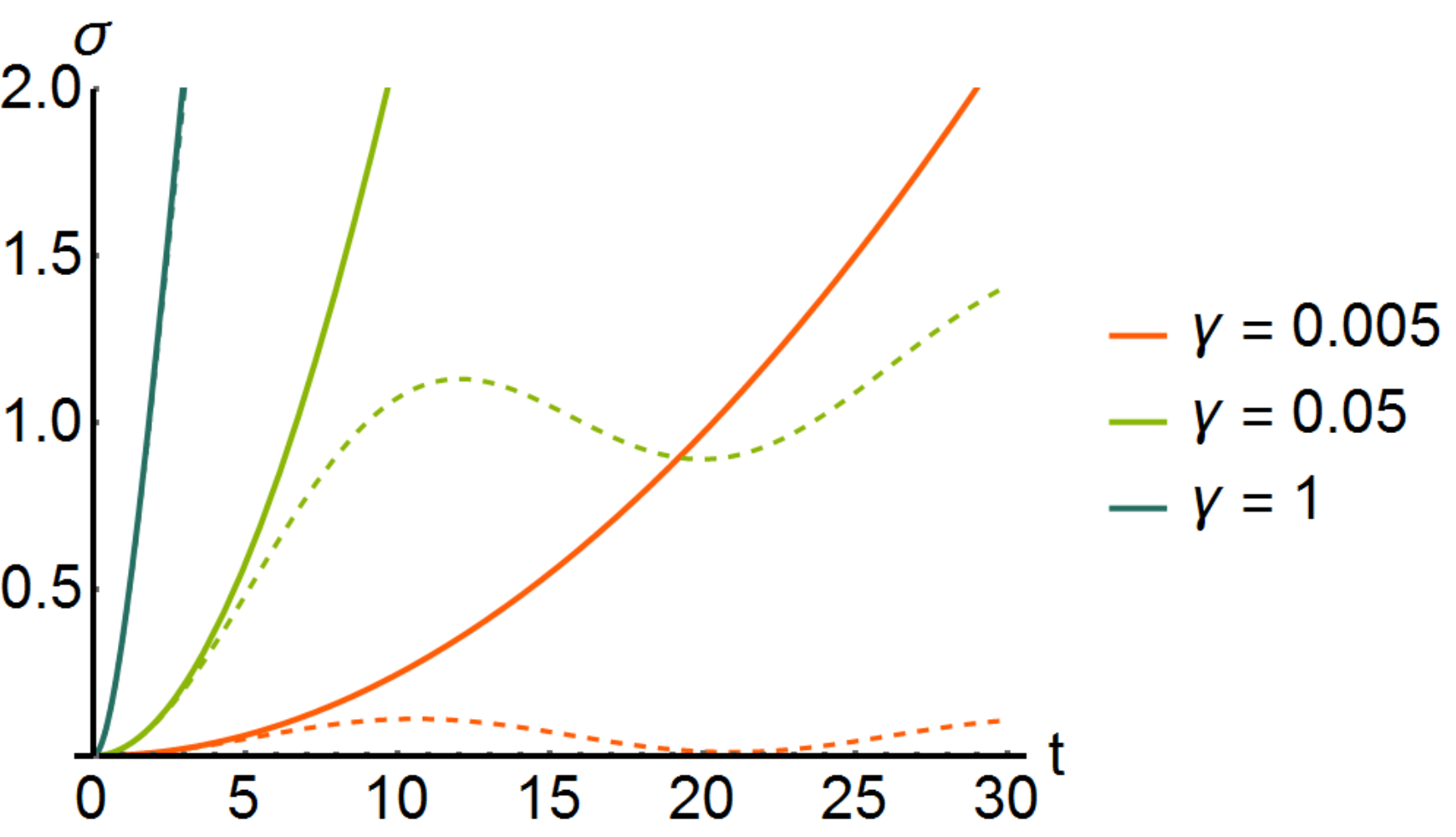}
\caption{\label{f:IOFsigma}(Color Online) 
\mgap{
Input-output fidelity as a function of the interaction time 
for a cat state with $\alpha=\sqrt{2}$ 
(upper-left panel) and a Fock state with $n=2$ (upper-right panel)
for different values of the memory parameter 
$\gamma$ and the detuning $\delta$ and for 
fixed coupling $\lambda = 1$. 
For comparison, the behaviour of the variance $\sigma$ 
(lower panel) in the same conditions.
In all the panels the orange curves are for $\gamma = 0.005$, 
green is for $\gamma = 0.05$, and blue refers to $\gamma = 1$ (blue). 
The solid curves correspond to zero detuning (resonant case) 
and the dashed ones are for $\delta=0.3$.
}} \end{figure}

The $F_{IO}$s for a cat state and a Fock state interacting 
with a classical environment are given by
\be
\label{fiocat}
F_{IO}^{(cat)}(t) = \frac{1+ 4 e^{2 | \alpha|^2}+ e^{4 | 
\alpha|^2}+ e^{\frac{4 | \alpha|^2}{1+ \sigma (t)}}+ 
e^{\frac{4 \sigma(t) | \alpha|^2}{1+ \sigma (t)}}}{2 [1
+ \sigma (t)]\left[1+ e^{2 | \alpha |^2}\right]^2} \,,
\ee
\begin{multline}
\label{fiofock}
F_{IO}^{(Fock)}(t) = \frac{1}{\sqrt{\pi} (1
+ \sigma(t))} \frac{\Gamma(n+ \frac12)}{\Gamma(n+1)}\times \\ 
\times {}_2 F_1\left(-n,\frac12,\frac12 -n, 
\left[\frac{1-\sigma(t)}{1+\sigma(t)}\right]^2\right)
\end{multline}
where ${}_2 F_1(a,b,c,;x)$ is a hypergeometric function.
The behaviour of the IO fidelities 
in Eqs. (\ref{fiocat}) and (\ref{fiofock}) as a function 
of the interaction time is reported in the upper 
panels of Fig. \ref{f:IOFsigma} \mgap{for $\alpha=\sqrt{2}$ and
$n=2$.}
\par
\mgap{For zero detuning, the IO fidelities show a monotonous 
behaviour with the memory parameter $\gamma$, which determines 
how fast the $F_{IO}$s decrease. This corresponds to a 
decoherence dynamics where the state evolves loosing memory of its 
initial conditions.  Conversely, in the presence of detuning the 
nonresonant curves detach from the respective resonant ones 
and $F_{IO}$s show a non monotonous behaviour where the 
state goes back to the initial preparation, at least partially.
Notice that the input-output fidelities of the cat and the Fock 
states shows a similar behavior, even though they are 
quantitatively different. In both cases, the
oscillating behaviour is present for values of $\gamma$ 
up to a threshold $\gamma^{*}$, which depends on the value of 
the other  parameters $\lambda$, and $\delta$ and {\em do not}
depend on the initial state, i.e. it represents a property of the
channel. For $\lambda=1$ and $\delta=0.3$ (see Fig. \ref{f:IOFsigma}) 
we have $\gamma^{*} \simeq 0.082$.} Loosely speaking, 
the existence of the threshold parameter $\gamma^{*}$ means that the
evolution is monotonous as far as the time correlations of the
environment are weak enough. Remarkably, the same kind of transition 
may be also seen in the time dependence of the variance $\sigma(t)$. As 
it is apparent from the lower panel of Fig. \ref{f:IOFsigma}, the presence 
of revivals in the behaviour of $\sigma$ also depends 
on the value of the memory parameter
$\gamma$ and it may be proven numerically that also the revivals disappear
when $\gamma \gtrsim \gamma^*$. Overall, this confirms  the backflow of
information and the ability of the input-output fidelity to capture
this feature of the quantum channel.
\section{Power-law process}
\label{s:pwl}
As mentioned in the introduction, the main conclusions of our analysis 
are qualitatively independent on the 
nature of the CSF used to model the environment. In order to
show this explicitly, and to briefly illustrate the 
quantitative effects of a differrent modelling, we report 
here the results obtained for a Gaussian process characterized
by a long range power-law autocorrelation
function of the form:
\be
\label{plaw}
K(t_1,t_2) = \frac{\beta-1}{2} 
\frac{\gamma \lambda}{(1+ \gamma |t_1-t_2|)^{\beta}}
\ee
where $\beta > 2$.
The attention will be focused on the nonclassical depth criterion, as it
is the most relevant one. The explicit form of $\sigma(t)$ for the
power-law process in the case of the resonant interaction ($\delta=0$)
is the following:
\be
\sigma (t) = \lambda t+ 
\lambda \frac{(1+ \gamma t)^{2-\beta}-1}{\gamma (\beta -2)}. 
\ee
\BINArev{This expression} can be approximated in some particular regimes to:
\begin{flalign}
\label{andam2}
\sigma(t) &\simeq \lambda t 
+ \frac{\lambda \gamma t^2}{(\beta -2) 
(1+ \gamma t)^{\beta}} &\quad  (\gamma \gg 1) \\
\sigma(t) &\simeq \frac{\lambda t^2}{2}  (\beta-1) \quad & (\gamma \ll 1).
\end{flalign}
As we can see from ($\ref{andam2}$), for $\gamma \to \infty$ the
nonclassical depth time approaches the Markovian limit 
$\sigma(t)\propto t$. Also for the power-law
process $\gamma$ plays the role of a memory parameter.  

\begin{figure}[h!]
\includegraphics[width=0.495 \columnwidth]{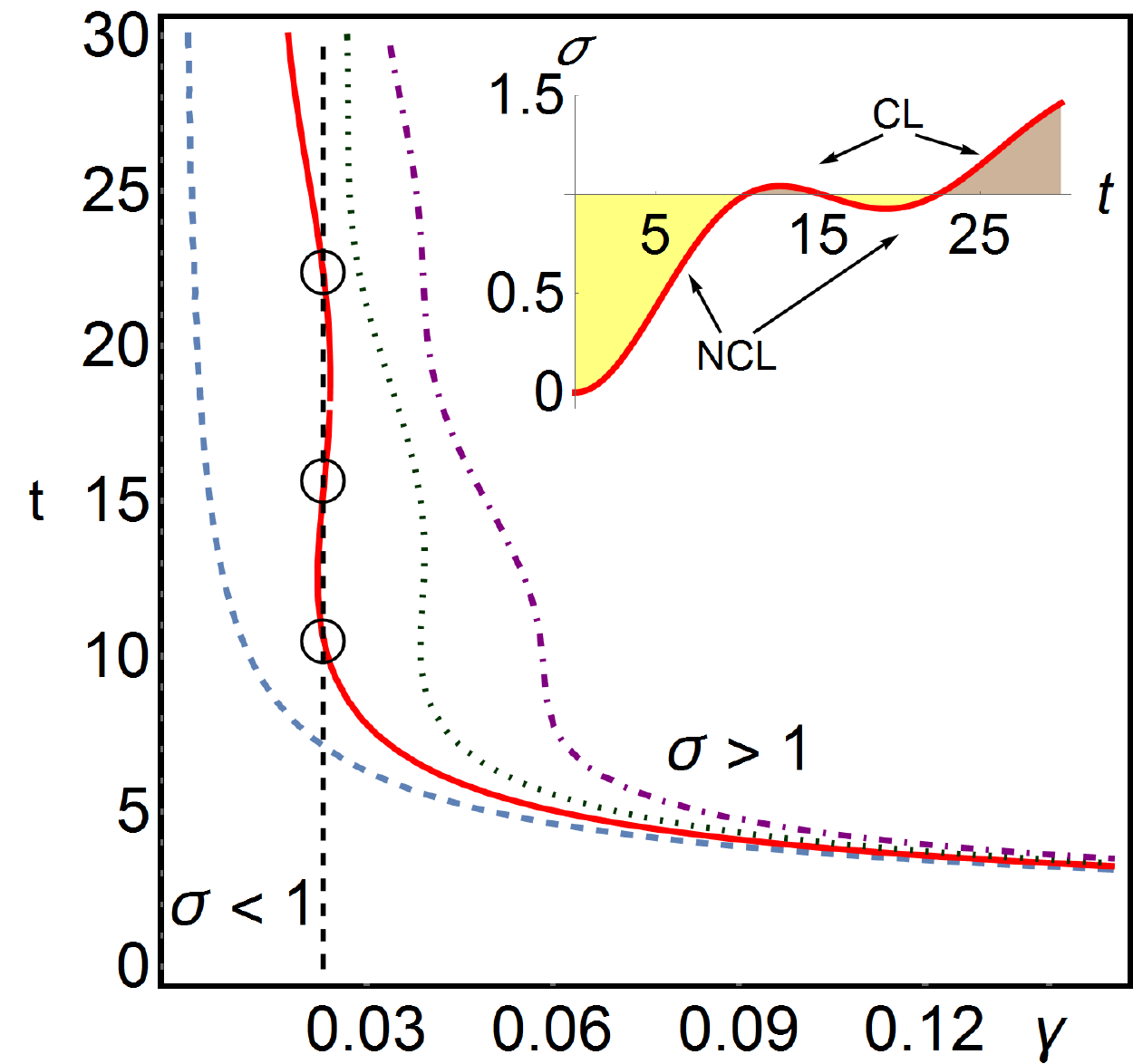}
\includegraphics[width=0.495 \columnwidth]{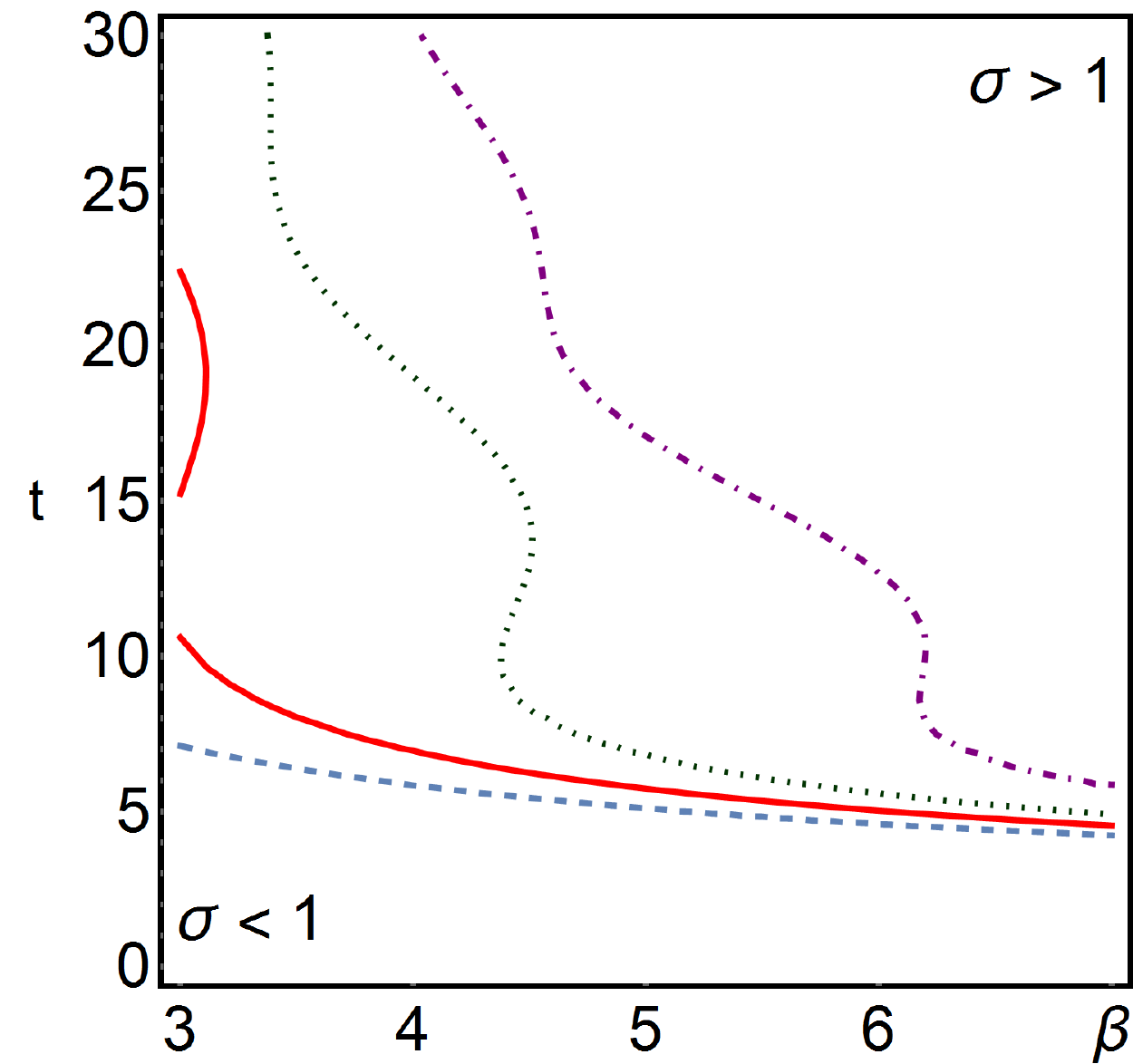}
\caption{(Color online) 
Dynamics of quantumness according to the nonclassical depth criterion
for a cat states evolving in classical environment with power-law
autocorrelation function.  In both panels, \BINArev{the dashed blue
curve represents the resonant case $\delta =0$, whereas solid red
($\delta =0.3$), dotted green ($\delta = 0.4$),  dot-dashed purple
($\delta =0.5$) curves refer to the off-resonance case}. Left panel:
nonclassical depth time $t_Q$ as a function of $\gamma$ in the case of a
Gaussian power-law process\BINArev{, for fixed $\beta=3$ and
$\lambda=1$}.  For $\gamma \gg 1$ the nonclassical depth time $t_Q$
approaches the Markovian limit independently of $\delta$. Sudden death
and birth of quantumness are highlighted by the circles along the dashed
black line at $\gamma=0.023$ \BINArev{and, correspondingly, in the
inset.} Right panel: nonclassical depth time $t_Q$ as a function of
$\beta$ \BINArev{and fixed $\gamma=0.023$}. } \label{f:plawalpha}
\end{figure}

In the nonresonant case, the analytic form of $\sigma(t)$ is
extremely complex and is not reported in this paper, whereas the results are 
explained in the following.
The presence of
sudden death and sudden birth of quantumness for the nonresonant interaction is shown, 
\mgap{for an initial
cat state}, in the left 
panel of Fig. \ref{f:plawalpha}, where, for fixed $\gamma$ and different
choices of the detuning parameter $\delta\neq 0$, we can see more than
one value of time $t_Q$ for which the nonclassical depth criterion is 
satisfied. In the right panel of Fig. \ref{f:plawalpha} we show 
the nonclassical depth time as a function of the parameter $\beta$ 
of  power-law autocorrelation function.
Furthermore, the presence of sudden death and birth of nonclassicality
depends not only on the particular combination of
parameters ($\delta, \gamma$), but also on the parameter $\beta$ itself. 
Actually, Fig. \ref{f:plawalpha} shows that nonclassicality revivals can
be also observed for the power-law process just like for the OU process,
and that this phenomenon is mostly due to the introduction of the detuning
parameter.
\section{Conclusions}
\label{s:out}
In this paper we have investigated the quantum-to-classical transition
for an harmonic oscillator initially prepared in a maximally
nonclassical state and then interacting with a classical fluctuating
field. As a first result, we have shown that modeling the environment by
means of classical stochastic fields allows to properly describe the
decoherence process in the presence of memory effects, without resorting
to approximated quantum master equations. In particular, we have been
able to introduce non-Markovian effects in a controlled way and to
recover the Markovian behavior by a suitable set of limiting values of
the parameters. 
\par
Our results show that the presence of classical memory in the environment 
strongly influences the decoherence time, increasing 
the survival time of nonclassicality  and leading to dynamical 
sudden death and birth of quantumness.
In particular, when the environment spectrum contains 
the natural frequency of the oscillator we observe an increase
of the survival time compared to the Markovian case whereas,
in the presence of a detuning, we see the occurrence of sudden death 
and sudden birth of quantumness, as indicated by collapses and 
revivals of nonclassicality. In order to address this phenomena
quantitatively, we have
analyzed the behavior of four different criteria introduced to
witness nonclassicality, also relating them to experimentally observable
quantities. All
this quantifiers agree in describing the nontrivial decoherence
process and the revivals of nonclassicality, thus supporting the 
validity of our model and the 
main conclusions of our analysis, which may be summarized as follows:
i) classical 
memory effects increase the survival time of quantumness; ii) a
detuning between the natural frequency of the system and the 
central frequency of the environment produces revivals of quantumness.
\begin{acknowledgments}
This work has been supported by MIUR (FIRB ``LiCHIS'' RBFR10YQ3H).
MGAP thanks C. Benedetti, M. G. Genoni and M. A. C. Rossi for
discussions.
\end{acknowledgments}
\begin{appendix}
\section{}
The $s$-ordered characteristic function of the cat state is given by
\begin{align}
\chi_s&[{\rm cat}] (\mu) = \frac{2}{\mathcal{N}}\,e^{-\frac{1}{2}(1-s)|\mu|^2} 
\notag \\ & \times \left[ \cos \left(2 \operatorname{Im} \mu \alpha^{*}
\right)+ e^{-2|\alpha|^2} \cosh \left( 2\operatorname{Re}\mu \alpha^{*}
\right) \right]
\end{align}
The $s$-ordered Wigner function of the cat state is given by
\begin{align}
W_s[&{\rm cat}] (\beta) = \frac{2\,e^{-\frac{2 |\beta|^2}{1-s}}}{\mathcal{N} 
\pi (1-s)} \notag \\  \times& \Bigg[ \exp \left \{ \frac{2 s |\alpha|^2}{1-s}
\right \} \cos \left( \frac{4}{1-s} \operatorname{Re}\beta \alpha^{*}\right ) 
\nonumber\\ & + \exp \left \{ -\frac{2 |\alpha|^2}{1-s}\right \} 
\cosh \left ( \frac{4}{1-s} \operatorname{Im}\beta \alpha^{*}\right )
\Bigg ]\,.
\end{align}
The matrix elements of the Schr\"odinger state in the Fock states basis are given by
\begin{align}
\rho_{n,m}=\frac{1}{\mathcal{N}} e^{-|\alpha|^2}\frac{\alpha^n(\alpha^{*})^m}{\sqrt{n!m!}}[1+(-1)^n][1+(-1)^m].
\end{align}
The $s$-ordered characteristic function of the Fock state is given by
\begin{align}
\chi_s[n] (\mu) = \exp \left \{- \frac{(1-s) |\mu|^2}{2} 
L_n (|\mu|^2) \right \}\,,
\end{align}
where $L_n (x)$ is the Laguerre polynomial of order $n$.
\par
The $s$-ordered Wigner function of the Fock state is given by
\begin{align}
W_s&[n] (\beta) = (-1)^n \frac{2}{\pi (1-s)}
\left(\frac{1+s}{1-s}\right)^n \notag \\ 
& \times \exp \left \{-\frac{2|\beta|^2}{1-s} \right \} 
\,L_n \left( \frac{4 |\beta|^2}{1-s^2}\right)\,.
\end{align}
\end{appendix}

\end{document}